\documentclass[a4paper,11pt]{article}
\pdfoutput=1 

\usepackage{jheppub} 

\usepackage[T1]{fontenc}
\usepackage{multirow}

\allowdisplaybreaks
\definecolor{nicered}{rgb}{0.7,0.1,0.1}
\definecolor{nicegreen}{rgb}{0.1,0.5,0.1}
\definecolor{niceblue}{rgb}{0.0,0.1,0.7}
\hypersetup{colorlinks,citecolor=niceblue,linkcolor=niceblue,urlcolor=niceblue}

\definecolor{color1}{rgb}{1.00,0.57,0.35}
\definecolor{color2}{rgb}{0.94,0.35,0.42}
\definecolor{color3}{rgb}{1.00,0.99,0.37}
\definecolor{color4}{rgb}{0.44,0.41,0.66}
\definecolor{color5}{rgb}{0.93,0.34,0.66}
\definecolor{color6}{rgb}{0.69,0.81,0.35}
\definecolor{color7}{rgb}{0.65,0.51,0.72}
\definecolor{color8}{rgb}{0.73,0.62,0.49}
\definecolor{color9}{rgb}{0.43,0.62,0.59}
\definecolor{color10}{rgb}{0.34,0.67,0.95}

\usepackage[normalem]{ulem}
\usepackage{caption}
\usepackage{subcaption}
\usepackage{fancyvrb} 

\def \bm#1{\mbox{\boldmath$#1$\unboldmath}}
\def \beq{\begin{equation}}
\def \eeq{\end{equation}}
\def \bea{\begin{eqnarray}}
\def \eea{\end{eqnarray}}

\title{Novel collider signatures \\ in the type-I~2HDM+$\bm{a}$ model}

\author[a,b]{Spyros Argyropoulos,}
\author[c]{Ulrich Haisch}
\author[a]{and Ilia Kalaitzidou}

\affiliation[a]{Physikalisches Institut, Universit{\"a}t Freiburg,\\ Hermann-Herder Str. 3a, 79104 Freiburg, Germany}
\affiliation[b]{Department of Physics, Aristotle University of Thessaloniki, \\ 54124 Thessaloniki, Greece}
\affiliation[c]{Max Planck Institute for Physics, \\ F{\"o}hringer Ring 6, 80805 M{\"u}nchen, Germany}

\preprint{MPP-2024-66}

\emailAdd{spyros.argyropoulos@cern.ch,haisch@mpp.mpg.de,ilia.kalaitzidou@cern.ch}

\abstract{The 2HDM+$a$ model is one of the main models used in the interpretations of dark matter searches at the LHC. So far, all the 2HDM+$a$ benchmarks considered by the ATLAS and CMS experiments are limited to a type-II Yukawa sector, in which the Higgs bosons~$A$, $H$, and~$H^\pm$ are all constrained to be mass-degenerate and heavier than around $600 \, {\rm GeV}$. In this~work, we present the first detailed study of 2HDM+$a$ models with a type-I Yukawa sector, which, for moderate values of $\tan\beta$, lift the constraints from flavour physics, allowing the extra Higgs bosons to be even lighter than the $125 \, {\rm GeV}$ Higgs boson discovered at the~LHC. We~discuss several benchmarks where the $A$, $H$, and~$H^\pm$ states are not necessarily mass-degenerate and the signatures that arise in these models, some of which have not yet been explored at the~LHC. We present the dominant channels in the studied benchmarks and the expected sensitivity in Run~2 data using truth-level analyses and discuss potential improvements in the experimental searches for~Run~3.}

\begin{document} 
\maketitle
\flushbottom

\section{Introduction}
\label{sec:introduction}

It has long been posited that the Higgs sector of the Standard Model~(SM) or extensions thereof might be connected to the dark sector, given the fact that Lorentz and gauge invariant Higgs bilinears can accommodate both renormalisable and unrenormalisable interactions to dark-state operators that form an SM gauge singlet~\cite{Silveira:1985rk,Veltman:1989vw,Schabinger:2005ei,Patt:2006fw,OConnell:2006rsp,Kim:2006af,Barger:2007im,Kanemura:2010sh,Djouadi:2011aa,Mambrini:2011ik,Djouadi:2012zc,Alanne:2017oqj,Arcadi:2017kky,Balkin:2018tma,Ruhdorfer:2019utl,Arcadi:2019lka,Das:2020ozo,Arcadi:2021mag,Lebedev:2021xey,Haisch:2022rkm,Haisch:2023aiz}. The discovery of an SM-like Higgs boson by the ATLAS and CMS collaborations~\cite{ATLAS:2012yve,CMS:2012qbp} has in consequence paved the way towards the full exploration of such Higgs-portal models, an endeavour which holds a central place in the search programmes of the LHC experiments.

Searches for dark matter~(DM) at the LHC (see for example~\cite{Kahlhoefer:2017dnp,Boveia:2018yeb,Argyropoulos:2021sav,Arcadi:2024ukq} for recent reviews) have initially employed simplified models with a single SM-singlet mediator as proposed by the LHC Dark Matter Working Group~(LHC~DM~WG)~\cite{Abdallah:2015ter,Abercrombie:2015wmb}. Later investigations showed that the interactions between the DM mediator and the SM fermions are not invariant under the $SU(2)_L\times U(1)_Y$ gauge symmetry, a feature which leads to unitarity violation at high energies~\cite{Bell:2015sza,Kahlhoefer:2015bea,Haisch:2016usn,Englert:2016joy,Haisch:2018bby}. The latter can be remedied either by introducing an extra scalar singlet that mixes with the SM Higgs~\cite{Kim:2008pp,Kim:2009ke,Baek:2011aa,Lopez-Honorez:2012tov,Baek:2012uj,Fairbairn:2013uta,Carpenter:2013xra}, a possibility which is however severely constrained by the recent measurements of the SM Higgs couplings~\cite{ATLAS:2022vkf,ATLAS-CONF-2023-052,CMS:2022dwd}, or by extending the SM Higgs sector with additional doublets~\cite{Deshpande:1977rw,Barbieri:2006dq,Cao:2007rm,Dolle:2009ft,Miao:2010rg,Gustafsson:2012aj,Belanger:2015kga,Ilnicka:2015jba,Poulose:2016lvz,Datta:2016nfz,Hashemi:2016wup,Belyaev:2016lok,Dutta:2017lny,Wan:2018eaz,Kalinowski:2018ylg,Kalinowski:2018kdn,Dercks:2018wch,Kalinowski:2020rmb}. The two-Higgs-doublet model with an additional pseudoscalar~\cite{Ipek:2014gua,No:2015xqa,Goncalves:2016iyg,Bauer:2017ota}, hereafter referred to as the 2HDM+$a$ model, constitutes the simplest gauge-invariant and renormalisable extension of the simplified pseudoscalar DM model benchmarked in~\cite{Abdallah:2015ter,Abercrombie:2015wmb}, admitting a host of signatures for DM direct and indirect detection experiments, as well as for collider phenomenology. It has therefore evolved into one of the pillars of the LHC DM search programme~\cite{Pani:2017qyd,Tunney:2017yfp,Arcadi:2017wqi,LHCDarkMatterWorkingGroup:2018ufk,CMS:2018zjv,Boveia:2018yeb,ATLAS:2019wdu,Arcadi:2020gge,CMS:2020ulv,ATLAS:2020yzc,ATLAS:2021jbf,ATLAS:2021shl,ATLAS:2021gcn,Robens:2021lov,Argyropoulos:2021sav,CMS:2022sfl,Argyropoulos:2022ezr,Arcadi:2022dmt,ATLAS:2022znu,Arcadi:2022lpp,Haisch:2023rqs,ATLAS:2023ofo,ATLAS:2023ild,ATLAS:2023rvb,Arcadi:2024ukq}.

All the searches for the 2HDM+$a$ model at the LHC performed so far have considered only a Yukawa sector of type~II and a fully degenerate mass spectrum for the beyond the SM~(BSM) Higgs bosons. While these assumptions are well-motivated in view of electroweak~(EW) precision observables~\cite{Haller:2018nnx} and measurements of the couplings of the $125 \, {\rm GeV}$ Higgs boson~\cite{ATLAS:2024lyh}, they limit the range of experimental signatures that could emerge from an extended Higgs sector. In this paper, we show that the 2HDM+$a$ model with a moderately fermiophobic Yukawa sector of type~I can accommodate a large region of parameter space featuring non-degenerate BSM Higgs bosons, giving rise to experimental signatures that remain completely unexplored. Furthermore, we note that in the type-I~2HDM+$a$ model, larger departures from the alignment limit are allowed~\cite{ATLAS:2024lyh}, which can also lead to new experimental signatures (see~e.g.~\cite{Aguilar-Saavedra:2022xrb}). The goal of this paper is thus twofold: $(i)$ to establish a first set of concrete benchmark points that are theoretically consistent and fulfil the constraints imposed by flavour experiments as well as LEP and LHC measurements, and $(ii)$ to provide a first exploration of the experimental signatures that can arise, highlighting the complementarity of different searches in distinct regions of the parameter space. Since~the collider phenomenology of the proposed type-I~2HDM+$a$ benchmarks turns out to be rich and novel, we recommend that the proposed scenarios be considered as potential new benchmarks for the ATLAS and CMS DM searches in Run~3 of the LHC.

This article is organised as follows: Section~\ref{sec:2hdma} provides a general overview of the 2HDM+$a$ model, while a comprehensive review of the experimental and theoretical constraints that have to be satisfied to make a given type-I~2HDM+$a$ model phenomenologically viable is given~in~Section~\ref{sec:constraints}. In Section~\ref{sec:decaypattern}, we study the partial decay widths and branching ratios of the BSM spin-$0$ states. This analysis enables us to determine the dominant collider signature expected in the type-I 2HDM+$a$ model with a non-degenerate BSM Higgs sector. The~choices of the model parameters for the new type-I~2HDM+$a$ benchmarks that we propose are spelled out in~Section~\ref{sec:benchmarks}. In~Section~\ref{sec:results}, we determine the expected sensitivity of the dominant missing transverse energy ($E_{T,\rm miss}$) and non-$E_{T,\rm miss}$ channels in four of the studied benchmarks. We also provide a very brief discussion of the behaviour of the relic density and the DM direct detection limits in~Section~\ref{sec:relicdensity}. Section~\ref{sec:conclusions}~summarises the main findings of our work and presents an outlook. Additional material is relegated to the~appendices.

\section{Basics of the 2HDM+$\bm{a}$ model}
\label{sec:2hdma}

In this section, we provide a concise overview of the 2HDM+$a$ model, directing readers interested in more details to the earlier articles~\cite{Bauer:2017ota,LHCDarkMatterWorkingGroup:2018ufk,Robens:2021lov,Argyropoulos:2021sav,Argyropoulos:2022ezr,Haisch:2023rqs}. We write the 2HDM~part of the tree-level scalar potential of the 2HDM+$a$ model in the following way: 
\beq \label{eq:VH}
\begin{split}
V_H & =\mu_1H_1^{\dag}H_1+\mu_2H_2^{\dag}H_2+\left(\mu_3H_1^{\dag}H_2+\mathrm{h.c.}\right)+\lambda_1\left(H_1^{\dag}H_1\right)^2+\lambda_2\left(H_2^{\dag}H_2\right)^2 \\[2mm]
 & \phantom{xx} +\lambda_3\left(H_1^{\dag}H_1\right)\left(H_2^{\dag}H_2\right)+\lambda_4\left(H_1^{\dag}H_2\right)\left(H_2^{\dag}H_1\right)+\left[\lambda_5\left(H_1^{\dag}H_2\right)^2+\mathrm{h.c.}\right] \,. 
\end{split} 
\eeq
Here a $\mathbb{Z}_2$ symmetry is enforced, wherein the two Higgs doublets $H_{1,2}$ undergo transformations as $H_1\to H_1$ and $H_2\to -H_2$, respectively. The $\mathbb{Z}_2$ symmetry serves as the minimal requirement to ensure the absence of flavour-changing neutral currents at the tree level. It is worth noting that all terms in~(\ref{eq:VH}) except $\mu_3H_1^{\dag}H_2+\mathrm{h.c.}$ adhere to this discrete symmetry. The vacuum expectation values (VEVs) of the Higgs doublets are given by $\langle H_{1,2} \rangle = (0, v_{1,2}/\sqrt{2})^T$ with $v = (v_1^2 + v_2^2)^{1/2}$. To address potential problems related to electric dipole moments, the additional assumption is made that the parameters $\mu_{1,2,3}$ and $\lambda_{1,2,3,4,5}$ appearing in $V_H$ are all real. Consequently, the CP eigenstates arising from the spontaneous symmetry breaking of~(\ref{eq:VH}) can be identified with the mass eigenstates. Specifically, two scalar states, denoted as $h$ and $H$, emerge along with one pseudoscalar particle, denoted as $A$, and two charged scalars, denoted as $H^\pm$.

Apart from~(\ref{eq:VH}), the tree-level scalar potential in the 2HDM+$a$ model encompasses the following two contributions:
\beq \label{eq:VHPVP}
\begin{split}
V_{HP} & = P \left ( i \hspace{0.125mm} b_P \hspace{0.25mm} H_1^\dagger H_2 + {\rm h.c.} \right ) + P^2 \left ( \lambda_{P1} \hspace{0.25mm} H_1^\dagger H_1 + \lambda_{P2} \hspace{0.25mm} H_2^\dagger H_2 \right ) \,, \\[2mm]
V_{P} & = \frac{1}{2} \hspace{0.25mm} m_P^2 \hspace{0.25mm} P^2 + \lambda_{P3} \hspace{0.25mm} P^4 \,.
\end{split}
\eeq
Here $P$ is an additional pseudoscalar mediator that transforms as $P \to P$ under the $\mathbb{Z}_2$~symmetry. The parameters $b_P$ and $\lambda_{P1, P2, P3}$ are chosen to be real, ensuring that~(\ref{eq:VHPVP}) conserves~CP. Notice~that the first term in $V_{HP}$ softly breaks the $\mathbb{Z}_2$~symmetry. Since $P$ does not acquire a VEV, the quartic self-coupling $\lambda_{P3}$ only gives rise to four-point interactions of the CP-odd spin-0 states and leaves their masses, mixings, trilinear couplings and decay widths unchanged at tree level. As a result, it has no effect on any of the observables discussed in~Section~\ref{sec:results}. The $P^4$ term is, however, essential for the bounded-from-below (BFB) conditions of the 2HDM+$a$ model, as will be explained in~Section~\ref{sec:theoryconstraints}.

If the Dirac DM field $\chi$ is taken to transform under the $\mathbb{Z}_2$~symmetry as $\chi \to -\chi$, the only renormalisable DM-mediator coupling that is allowed by this symmetry is
\beq \label{eq:2HDMaLchi}
{\cal L}_\chi = - i \hspace{0.125mm} y_\chi \hspace{0.125mm} P \hspace{0.25mm} \bar \chi \gamma_5 \chi \,,
\eeq
where it is assumed again that the dark-sector Yukawa coupling $y_\chi$ is real, making all BSM~interactions in the 2HDM+$a$ model~CP conserving. 

Taking into account the mass $m_\chi$ of the DM particle, the 2HDM+$a$ model introduces~15~new parameters in addition to the SM ones. Upon rotation to the mass eigenstate basis, these parameters can be exchanged for seven physical masses, three mixing angles, and five~couplings:
\beq \label{eq:2HDMainput}
\begin{Bmatrix}
\mu_1, \hspace{0.125mm} \mu_2, \hspace{0.125mm} \mu_3, \hspace{0.125mm} b_P, \hspace{0.125mm} m_P, \hspace{0.125mm} m_{\chi},\\
y_{\chi}, \hspace{0.125mm} \lambda_1, \hspace{0.125mm} \lambda_2, \hspace{0.125mm} \lambda_3, \hspace{0.125mm} \lambda_4, \hspace{0.125mm} \lambda_5,\\
\lambda_{P1}, \hspace{0.125mm} \lambda_{P2}, \hspace{0.125mm} \lambda_{P3}
\end{Bmatrix} \ \
\Longleftrightarrow
\ \
\begin{Bmatrix}
v, \hspace{0.125mm} m_h, \hspace{0.125mm} m_A, \hspace{0.125mm} m_H, \hspace{0.125mm} m_{H^{\pm}}, \hspace{0.125mm} m_a, \hspace{0.125mm} m_{\chi},\\
 \tan\beta, \hspace{0.125mm} \cos \left (\beta-\alpha \right ), \hspace{0.125mm} \sin\theta,\\
y_{\chi}, \hspace{0.125mm} \lambda_3, \hspace{0.125mm} \lambda_{P1}, \hspace{0.125mm} \lambda_{P2}, \hspace{0.125mm} \lambda_{P3}
\end{Bmatrix} \,.
\eeq
Here $\tan\beta=v_2/v_1$ encodes the ratio of the VEVs~$v_{1,2}$ of the two Higgs doublets, while the angle $\alpha$ describes the mixing of the two neutral~CP-even weak spin-$0$ eigenstates. In~this work we employ the conventions of~\cite{Bauer:2017ota}, in which the couplings of the~CP-even Higgs bosons to the EW gauge bosons are given as
\beq \label{eq:HVVcoupling}
\mathcal{L}\supset \Big[ \sin \left (\beta-\alpha \right ) h+\cos \left (\beta-\alpha \right )H \Big ] \, \left(\frac{2 m_W^2}{v} \, W_{\mu}^+W^{-\mu}+\frac{m_Z^2}{v} \, Z_{\mu}Z^{\mu}\right) \,,
\eeq
with $m_W$ and $m_Z$ denoting the mass of the $W$ and $Z$ boson, respectively. Notice that via~(\ref{eq:HVVcoupling}) the state $h$ is identified with the $125 \, {\rm GeV}$ Higgs boson in the so-called alignment limit $\cos \left (\beta - \alpha \right ) = 0$, while $H$ indicates the BSM~CP-even Higgs boson, independently of the mass hierarchy of the two~CP-even Higgs bosons. The variable $\sin\theta$ in~(\ref{eq:2HDMainput}) furthermore encodes the mixing of the two CP-odd weak spin-$0$ eigenstates and the additional pseudoscalar mediator $a$ is mostly composed of $P$ for $\sin \theta \simeq 0$. The parameters appearing on the right-hand side of~(\ref{eq:2HDMainput}) are used as input in the analyses of the 2HDM+$a$~model. Since the~VEV~$v \simeq 246 \, {\rm GeV}$ and the mass $m_h \simeq 125 \, {\rm GeV}$ of the SM-like Higgs boson are already fixed by observations there are in total 13 input parameters.

\section{Experimental and theoretical constraints}
\label{sec:constraints}

In this section, we examine the constraints on the parameters~(\ref{eq:2HDMainput}) of the type-I~2HDM+$a$ model that direct and indirect searches for spin-$0$ states provide. Theoretical restrictions on the parameter space are also examined. Our discussion offers a first insight into the kind of 2HDM+$a$ models of type~I that can exhibit a distinctive LHC phenomenology.

\subsection{Higgs-boson physics}
\label{sec:higgsphysics}

The LHC measurements of the $125 \, {\rm GeV}$ Higgs-boson couplings~\cite{ATLAS:2022vkf,ATLAS-CONF-2023-052,CMS:2022dwd} provide tight constraints on $\cos \left (\beta-\alpha \right )$. These are particularly strong for the type-II, lepton-specific, and flipped 2HDM models --- see~\cite{ATLAS:2023qur} for a very recent detailed discussion --- and have motivated the choice $\cos \left (\beta-\alpha \right )=0$ in all 2HDM+$a$ benchmarks~\cite{LHCDarkMatterWorkingGroup:2018ufk} considered so far in the existing ATLAS and CMS interpretations~\cite{CMS:2018zjv,ATLAS:2019wdu,CMS:2020ulv,ATLAS:2020yzc,ATLAS:2021jbf,ATLAS:2021shl,ATLAS:2021gcn,CMS:2022sfl,ATLAS:2022znu,ATLAS:2023ofo,ATLAS:2023ild,ATLAS:2023rvb}. We~note that in the type-I benchmarks proposed below deviations from the alignment limit are allowed, particularly in the parameter space with $\tan\beta\gtrsim 2$. Small deviations from alignment would open up additional experimental signatures, such as the decays $A \to Zh$, $H \to ZZ$, $H \to W^+W^-$, $H \to hh$ and others, which are absent when $\cos \left (\beta-\alpha \right )=0$~\cite{Gunion:2002zf,Craig:2013hca,Kling:2016opi}. To simplify matters, we set aside this possibility and, from this point onward, exclusively concentrate on the exact alignment limit.

If the pseudoscalar $a$ is sufficiently light, the $125 \, {\rm GeV}$ Higgs boson discovered at the LHC can decay into a pair of such CP-odd states. If in addition $m_\chi < m_a/2$ the decay $h \to aa$ followed by $a \to \chi \bar \chi$ will lead to an invisible Higgs decay signal. The latest searches for invisible and undetected decays of the Higgs boson~\cite{CMS:2022qva,ATLAS:2024lyh} impose a lower limit of $m_a \gtrsim 100 \, {\rm GeV}$ on the pseudoscalar mass~\cite{Bauer:2017ota} unless the trilinear coupling $g_{haa}$ is tuned such that $\Gamma \left ( h \to a a \right ) \lesssim 1 \, {\rm MeV}$~\cite{Argyropoulos:2022ezr,Haisch:2023rqs}. Notice that the latter bound on $m_a$ is stronger than $m_a > m_h/2$, that one would naively expect, due to the off-shell contributions to the four-body decays of the form $h \to aa \to f \bar f \chi \hspace{0.5mm} \bar \chi$ with $f$ denoting an SM fermion. Since in this work we are mainly interested in mono-$X$ signatures that are triggered by the decay $a \to \chi \bar \chi$ we will consider only benchmarks with the mass hierarchy $m_\chi \ll 100 \, {\rm GeV} \lesssim m_a$. In this case the bounds from invisible and undetected decays of the $125 \, {\rm GeV}$ Higgs boson are automatically satisfied without tuning. 

We add that when the decay $a \to \chi \bar \chi$ is kinematically allowed,~i.e.~for $m_a > 2 m_{\chi}$, the DM relic density is in general overabundant in the 2HDM+$a$ model~\cite{LHCDarkMatterWorkingGroup:2018ufk,Arcadi:2024ukq,Argyropoulos:2022ezr}. Since~the~$E_{T,\rm miss}$ signatures studied below are generated by the $a\to \chi\bar{\chi}$ decay, this implies that finding benchmarks where $E_{T,\rm miss}$ signals are dominant while simultaneously producing the correct relic density is challenging. Further details on the DM phenomenology in the type-I~2HDM+$a$ model can be found in~Section~\ref{sec:relicdensity}.

\subsection{Flavour physics and BSM Higgs searches}
\label{sec:flavourphysics}

While flavour observables such as $B \to X_s \gamma$ and $B_s \to \mu^+ \mu^-$ place tight constraints on the mass of the charged Higgs boson (i.e.~$m_{H^{\pm}}\gtrsim 600$ GeV) in the case of Yukawa sector of type~II, these constraints are generally much weaker and essentially absent for $\tan\beta\gtrsim 3$ in the case of 2HDM models of type~I~\cite{Fox:2017uwr,Haisch:2017gql,Bauer:2017ota,LHCDarkMatterWorkingGroup:2018ufk,Robens:2021lov,Haller:2018nnx}. To accommodate a lighter spectrum of BSM spin-$0$ states, we diverge from~\cite{Bauer:2017ota,LHCDarkMatterWorkingGroup:2018ufk} and take in this article the Yukawa sector of the 2HDM+$a$ model to be of type~I. With this choice and working in the alignment limit, the couplings between the $H$, $A$, $a$ states and the SM fermions take the~form: 
\beq \label{eq:Lag2HDMfermions}
{\cal L} \supset  \frac{g_{H \bar{f} f}}{\sqrt{2}} \, H \bar f f - \frac{i g_{A \bar{f} f}}{\sqrt{2}} \, A \bar f \gamma_5 f  + \frac{i g_{a \bar{f} f}}{\sqrt{2}} \, a \bar f \gamma_5 f \,, 
\eeq
with
\beq \label{eq:2HDMfermions}
g_{H \bar{f} f} = y_f \hspace{0.25mm} \cot \beta \,, \qquad g_{A \bar{f} f} = \eta_f \hspace{0.5mm} y_f \hspace{0.25mm} \cot \beta \hspace{0.125mm} \cos \theta \,, \qquad g_{a \bar{f} f} = \eta_f \hspace{0.5mm} y_f \hspace{0.25mm} \cot \beta \hspace{0.125mm} \sin \theta \,.
\eeq 
Here $y_f = \sqrt{2} \hspace{0.125mm} m_f/v$ with $f=u,d,\ell$ denotes the SM Yukawa couplings, $m_f$ are the respective fermion masses and $\eta_u = 1$ while $\eta_d = \eta_\ell = -1$. Notice that all couplings in~(\ref{eq:2HDMfermions}) are suppressed for large $\tan \beta$. The rates of $H$, $A$, and $a$ production in gluon-gluon-fusion~(ggF), $b \bar b$-fusion or in association with $t \bar t$ or $b \bar b$ pairs can therefore be rendered unobservably small by taking $\tan \beta \gg 1$, even if the BSM Higgs bosons are light. Since the couplings of the charged Higgs boson to SM fermions are also $\tan \beta$ suppressed, the existing LHC limits on $H^\pm$ production from processes like $pp \to tb H^\pm$~\cite{CMS:2019rlz,ATLAS:2021upq} can also be avoided by making $\tan \beta$ sufficiently~large. As a result, the constraints arising from direct searches for BSM Higgs bosons are generally weaker in type-I than in type-II~2HDM models, in particular for $\tan \beta$~values~of~a~few. We will come back to this point in~Section~\ref{sec:results}. 

The possibility of a light BSM Higgs sector has regained interest due to several small excesses appearing in recent LHC searches --- see for example Section~2.9 of~\cite{Crivellin:2023zui} for a brief summary of the observed anomalies. The most well-known deviations are the hints of a~$\gamma \gamma$~resonance at around $95 \, {\rm GeV}$ reported by both ATLAS and CMS~\cite{CMS:2018cyk,ATLAS-CONF-2023-035,CMS-PAS-HIG-20-002}. The~excess~at $95 \, {\rm GeV}$ lies in a similar mass range as the lasting anomaly observed by LEP in the $e^+ e^- \to Z H \to \mu^+ \mu^- b\bar{b}$ channel~\cite{LEPWorkingGroupforHiggsbosonsearches:2003ing} and also coincides with the excess at about $100 \, {\rm GeV}$ reported by CMS~\cite{CMS:2022goy} in the $\tau^+ \tau^-$ channel, if the resolution of the reconstructed $\tau^+ \tau^-$ mass is taken into account. While various BSM models with a 2HDM sector can address these excesses~\cite{Fox:2017uwr,Haisch:2017gql,Bhatia:2017ttp,Biekotter:2019kde,Heinemeyer:2021msz,Biekotter:2022jyr,Biekotter:2022abc,Banik:2023ecr,Biekotter:2023jld,Azevedo:2023zkg,Biekotter:2023oen,Belyaev:2023xnv,Aguilar-Saavedra:2023tql,Dutta:2023cig,Arcadi:2023smv}, in this article we do not restrict ourselves to benchmark points which could potentially accommodate some or all of the aforementioned anomalies. We~discuss instead the phenomenology of more generic type-I~2HDM+$a$ benchmarks with a light BSM Higgs spectrum, focusing on the intriguing signatures that can arise in such scenarios. Before proceeding further, it is, however, worth noting that in 2HDM model realisations aimed at addressing the $95 \, {\rm GeV}$ excess, it is expected that the $95 \, {\rm GeV}$ resonance  decays frequently into $b \bar b$ pairs. For example, the four type-I~2HDM benchmarks analysed in the paper~\cite{Haisch:2017gql} all yield ${\rm Br} \left ( H \to b \bar b \right ) \gtrsim {\rm Br} \left ( h \to b \bar b \right ) \simeq 58\%$. This observation provides clear motivation for LHC searches targeting light BSM spin-$0$ states in final states containing bottom quarks.

\subsection{EW precision observables}
\label{sec:EWPO}

One important feature of the 2HDM+$a$ benchmarks proposed in~\cite{LHCDarkMatterWorkingGroup:2018ufk} and subsequently employed in the existing LHC interpretations~\cite{CMS:2018zjv,ATLAS:2019wdu,CMS:2020ulv,ATLAS:2020yzc,ATLAS:2021jbf,ATLAS:2021shl,ATLAS:2021gcn,CMS:2022sfl,ATLAS:2022znu,ATLAS:2023ofo,ATLAS:2023ild,ATLAS:2023rvb} is the assumption that the three BSM spin-$0$ states are degenerate in mass,~i.e.~$m_A = m_H = m_{H^\pm}$. This~choice, along with $\cos \left ( \beta - \alpha \right ) = 0$, is phenomenologically motivated because in such cases, the constraints from EW precision measurements, especially the~$\rho$~parameter, are automatically satisfied. More~specifically, in the aligned 2HDM+$a$ model, the one-loop correction to the $\rho$ parameter takes the form 
\beq \label{eq:Deltarho}
\Delta \rho = \frac{1}{(4 \pi)^2} \frac{1}{v^2} \Big [ \cos^2 \theta \, f (m_{H^\pm}^2,m_A^2,m_H^2) + \sin^2 \theta \, f (m_{H^\pm}^2,m_a^2,m_H^2) \Big] \,, 
\eeq
with 
\beq \label{eq:frho}
\begin{split}
f(m_1^2, m_2^2, m_3^2) & = m_1^2 - \frac{m_1^2 \hspace{0.25mm} m_2^2}{m_1^2 - m_2^2} \ln \left ( \frac{m_1^2}{m_2^2} \right ) \\[2mm] 
& \phantom{xx} - \frac{m_1^2 \hspace{0.25mm} m_3^2}{m_1^2 - m_3^2} \ln \left ( \frac{m_1^2}{m_3^2} \right ) + \frac{m_2^2 \hspace{0.25mm} m_3^2}{m_2^2 - m_3^2} \ln \left ( \frac{m_2^2}{m_3^2} \right ) \,. 
\end{split}
\eeq
The function~(\ref{eq:frho}) obeys
\beq \label{eq:fzero}
f (m_1^2, m_2^2, m_1^2) = 0 \,,
\eeq
which implies that the one-loop correction~(\ref{eq:Deltarho}) vanishes identically for $m_H = m_{H^\pm}$ irrespectively of the choices for $m_A$, $m_a$, and $\sin \theta$. The reason for the observed cancellation is that the tree-level scalar potential~$V_H+V_{HP}$~$\big($see~(\ref{eq:VH}) and~(\ref{eq:VHPVP})$\big)$ is custodially invariant for $m_H = m_{H^\pm}$~\cite{Bauer:2017ota}. 

The experimental $95\%$~confidence level~(CL) bound on the $\rho$ parameter extracted from a simultaneous determination of the Peskin-Takeuchi $S$, $T$, and $U$ parameters is~\cite{ParticleDataGroup:2022pth}:
\beq \label{eq:rhoconstraint}
\Delta \rho \in [-1.6, 2.0 ] \cdot 10^{-3} \,.
\eeq
It is important to realise that the EW fit of the PDG~\cite{ParticleDataGroup:2022pth} currently does not incorporate the CDF~II measurement of the $W$-boson mass~\cite{CDF:2022hxs}. This measurement shows significant tensions with earlier direct collider determinations as well as the SM prediction of $m_W$ obtained from precision EW data~\cite{Amoroso:2023pey}. As shown for instance in~\cite{Strumia:2022qkt,deBlas:2022hdk,Asadi:2022xiy}, interpreting the $m_W$~anomaly within the framework of non-zero $S$ and $T$ parameters but $U = 0$, suggests a positive shift in the $\rho$ parameter in the ballpark~of:
\beq \label{eq:rhoconstraintCDFII}
\Delta \rho \simeq 1.5 \cdot 10^{-3} \,.
\eeq
We add that the observed discrepancy in $m_W$ at CDF II can also be accommodated in the EW fit by employing $S \simeq T \simeq 0$ and $U \simeq 0.12$~\cite{deBlas:2022hdk,Asadi:2022xiy}, implying $\Delta \rho \simeq 0$. As~the~parameter~$U$ like $T$ violates custodial symmetry but corresponds to a dimension-eight and not dimension-six operator~\cite{Grinstein:1991cd}, it is however difficult to imagine a model that generates a large value of $U$ but simultaneously not of $T$. Solutions of the $m_W$ anomaly that lead to~(\ref{eq:rhoconstraintCDFII}) therefore have a more natural model-building interpretation. 

\begin{figure}[t!]
\begin{center}
\includegraphics[width=0.99\textwidth]{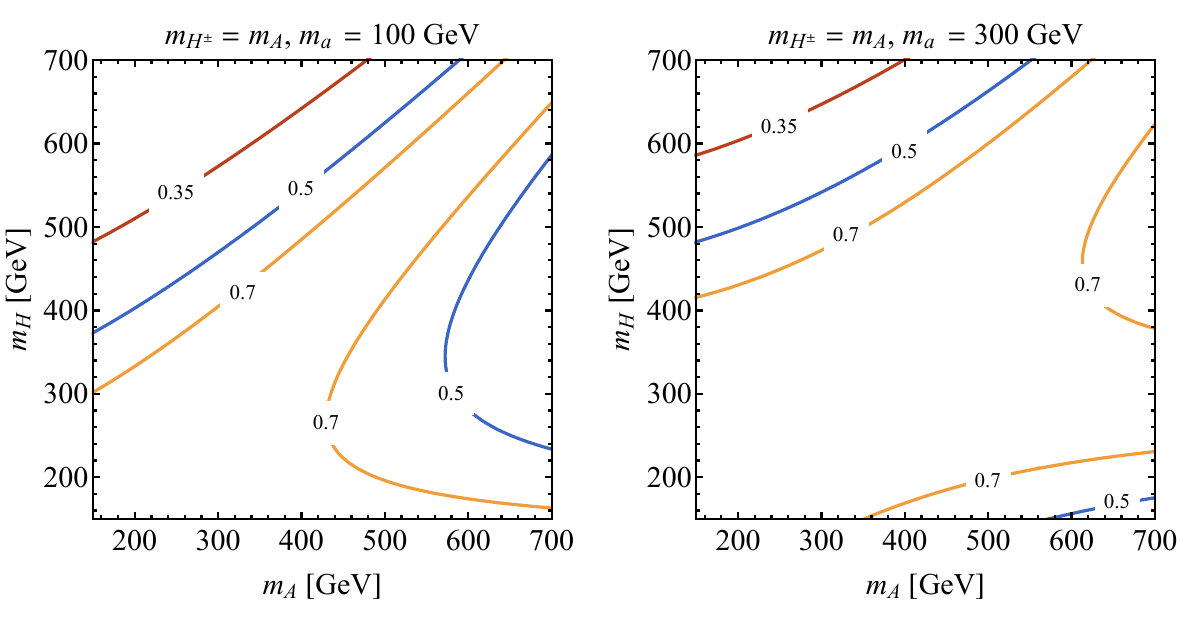}
\vspace{2mm}
\caption{\label{fig:rho} Values of $m_A$ and $m_H$ consistent with~(\ref{eq:rhoconstraint}) for three different choices of $\sin \theta$, as indicated by the contour labels. The parameter space above and to the right of the contours is excluded. The choices of $m_{H^\pm}$ and $m_a$ are given in the headlines of the plots and all shown constraints assume $\cos \left (\beta - \alpha \right ) = 0$. Consult the main text for additional~details.} 
\end{center}
\end{figure}

From~(\ref{eq:Deltarho}) one sees that in the aligned 2HDM+$a$ model, the breaking of the custodial symmetry is controlled by the mass splittings between the BSM Higgs bosons as well as the pseudoscalar mixing angle $\theta$. In fact, the stringent constraint on $\Delta \rho$ in general limits the size of mass non-degeneracy in the BSM Higgs sector. While the restriction to a fully or partially degenerate spectrum of $m_A$, $m_H$, and $m_{H^\pm}$ is thus motivated in view of~(\ref{eq:rhoconstraint}), it also has an important impact on the decay pattern of the BSM Higgs bosons, because it kinematically forbids decays such as $A \to H a$, $A\to ZH$, $A \to W^\mp H^\pm$, and others. As~will~be explained below, it is important to consider decays of this type, since they lead to experimental signatures which are not present in the degenerate type-II~2HDM+$a$ benchmarks. To~enlarge the space of possible LHC signatures, we will hence give up on the assumption of a mass-degenerate BSM Higgs spectrum, and study 2HDM+$a$ realisations in which the masses of the $A$, $H$, and $H^\pm$ states have splittings such that~(\ref{eq:rhoconstraint}) is satisfied. It~is also important to realise that in the case of a non-degenerate BSM Higgs sector, combining~(\ref{eq:Deltarho}) and~(\ref{eq:rhoconstraint}) in general also restricts the range of viable $\sin \theta$ values. Typically, for smaller pseudoscalar mixing angles $\theta$ larger mass splittings are allowed, and therefore in the type-I~2HDM+$a$ benchmark defined below, we will consider only $\sin \theta$ values significantly smaller than the maximal mixing case, i.e.~$\sin \theta = 1/\sqrt{2} \simeq 0.71$. In fact, it can be observed that the allowed parameter space is generically reduced for increasing~$\sin \theta$. Additionally, for the same $m_{H^\pm}$ and $\sin \theta$ values, the parameter space is smaller for $m_a = 100 \, {\rm GeV}$ compared to $m_a = 300 \, {\rm GeV}$. These features are illustrated in~Figure~\ref{fig:rho}, which shows the parameter space in the $m_A\hspace{0.125mm}$--$\hspace{0.5mm}m_H$ plane that is permitted by~(\ref{eq:rhoconstraint}) for $m_A = m_{H^\pm}$ and two different choices of $m_a$, assuming $\cos \left (\beta - \alpha \right ) = 0$.

\subsection{Theoretical restrictions}
\label{sec:theoryconstraints}

The requirement that the full scalar potential~$V_H+V_{HP}+V_P$ of the 2HDM+$a$ model is bounded from below~(BFB) restricts the possible choices of the spin-$0$ masses, the mixing angles, and the quartic couplings that appear on the right-hand side of~(\ref{eq:2HDMainput}). Under the assumption that $\lambda_{P1,P2} \geq 0$ and $\lambda_{P3} > 0$, the BFB tree-level conditions turn out to be identical to those in the pure 2HDM~\cite{Gunion:2002zf} when expressed in terms of $\lambda_{1,2,3,4,5}$.\footnote{Earlier versions of this article stated the BFB conditions for $\lambda_{P1,P2} \geq 0$ without including a $P^4$ term in $V_P$. As explained below, this is not sufficient for $\lambda_{P1} = \lambda_{P2} = 0$. In addition, the inequalities~(\ref{eq:BFB}) contained two misprints. We thank Daniel~Puigmart\'i~Molano for bringing both issues to our attention.} In the alignment limit, we find that the tree-level BFB conditions can be cast into the following inequalities:
\beq \label{eq:BFB}
\begin{split}
\lambda_3 & \geq c_1 \,, \qquad \lambda_3 \geq c_2 \,, \qquad \lambda_3 \geq - \sqrt{\left ( c_1 - \lambda_3 \right ) \left ( c_2 - \lambda_3 \right )} \,, \\[2mm]
& \hspace{8mm} c_3 - \big | c_4 - \lambda_3 \big | \geq - \sqrt{\left ( c_1 - \lambda_3 \right ) \left ( c_2 - \lambda_3 \right )} \,,
\end{split}
\eeq
with
\beq \label{eq:cis}
\begin{split}
c_1 & = \frac{1}{v^2} \hspace{0.5mm} \Big [ \hspace{0.5mm} m_h^2 \left ( 1 - \cot^2 \beta \right ) -2 m_H^2 + 2 m_{H^\pm}^2 \hspace{0.5mm} \Big ] \,, \\[2mm]
c_2 & = \frac{1}{v^2} \hspace{0.5mm} \Big [ \hspace{0.5mm} m_h^2 \left ( 1 - \tan^2 \beta \right ) -2 m_H^2 + 2 m_{H^\pm}^2 \hspace{0.5mm} \Big ] \,, \\[2mm]
c_3 & = \frac{1}{v^2} \hspace{0.5mm} \Big [ \hspace{0.5mm} m_h^2 - m_H^2 + m_A^2 \cos^2 \theta + m_a^2 \sin^2 \theta \hspace{0.5mm} \Big ] \,, \\[2mm]
c_4 & = \frac{1}{v^2} \hspace{0.5mm} \Big [ \hspace{0.5mm} m_h^2 - m_H^2 - m_A^2 \cos^2 \theta - m_a^2 \sin^2 \theta + 2 m_{H^\pm}^2 \hspace{0.5mm} \Big ] \,.
\end{split}
\eeq
Notice that in the limit of $m_A = m_H = m_{H^\pm}$ the four relations~(\ref{eq:BFB}) reduce to the two inequalities given in~Section~4.5 of~\cite{LHCDarkMatterWorkingGroup:2018ufk}. The BFB tree-level conditions, together with the perturbativity of the quartic coupling $\lambda_3$,~i.e.~the~requirement $|\lambda_3| \leq 4 \pi$, in general, preclude large mass splittings in the BSM Higgs sector. This is readily seen by considering the first inequality in~(\ref{eq:BFB}), which in the limit of large $\tan \beta$ reduces to $\lambda_3 \geq (m_h^2 -2 m_H^2 + 2 m_{H^\pm}^2)/v^2$. This implies that relatively large values of $\lambda_3$ are needed in order to maximise the parameter space that satisfies the BFB condition when a non-degenerate BSM spin-$0$ sector is considered. Since the tree-level BFB conditions are modified in more general 2HDM scalar sectors that include additional quartic couplings such as $\lambda_6 \hspace{0.25mm} \big( |H_1|^2 H_1^\dagger H_2 + {\rm h.c.} \big )$, we will treat~(\ref{eq:BFB}) only as a reference to indicate theoretically (dis)favoured parameter regions --- see the publications~\cite{Bauer:2017fsw,Haisch:2018djm} for more detailed discussions of this issue.

The requirement $\lambda_{P3} > 0$ is essential for $\lambda_{P1} = \lambda_{P2} = 0$, the choice adopted in the benchmarks studied below. In this case, the trilinear portal coupling $b_P$ contributes at the same order as the quartic couplings in~$V_H$ along field directions in which $P$ grows quadratically with the Higgs doublets, and for $\lambda_{P3} = 0$ it renders the potential unbounded from below in parts of the parameter space where~(\ref{eq:BFB}) is satisfied. A positive $\lambda_{P3}$ removes these directions, so that the potential is BFB if and only if~(\ref{eq:BFB}) holds. For small values of~$\lambda_{P3}$, minima with~$\langle P \rangle \neq 0$ that are deeper than the EW vacuum may, however, still exist. A detailed analysis of the global minimum of the scalar potential of the 2HDM+$a$ model is beyond the scope of this work.

In addition to identifying parameters that ensure vacuum stability, our numerical studies will also delineate regions in the parameter space where the total decay widths $\Gamma_{i}$ of all BSM spin-$0$ states with $i = A, H, H^\pm, a$ obey~$\Gamma_i/m_i \leq 30\%$. These criteria ensure that the total decay widths of the BSM spin-$0$ states are sufficiently small for the narrow width approximation~(NWA) to be applicable. While our Monte Carlo~(MC) studies consider effects from off-shell BSM Higgs production and decay, they do not account for potential modifications to the Higgs line shape~\cite{Seymour:1995np,Passarino:2010qk,Goria:2011wa}. The latter effects have been explored in~\cite{Anastasiou:2011pi,Anastasiou:2012hx}, revealing that for a heavy Higgs boson, different treatments of its propagator can lead to notable changes in inclusive production cross sections compared to the case of a Breit-Wigner with a fixed width, as employed in our article. Consequently, for parameter choices resulting in $\Gamma_i/m_i > 30\%$, the respective signal strengths exhibit some (difficult to quantify) model dependence related to the precise treatment of the BSM propagators. It is worth noting that to maintain small total decay widths $\Gamma_{i}$, parameter choices with small $\lambda_{P1,P2}$ are typically preferred. Since vacuum stability merely necessitates $\lambda_{P1,P2} \geq 0$ provided that $\lambda_{P3} > 0$, opting for $\lambda_{P1} = \lambda_{P2} = 0$ is hence a viable and simple choice from a phenomenological~perspective.

\section[Decay pattern of BSM spin-$\bm{0}$ states]{Decay pattern of BSM spin-$\bm{0}$ states}
\label{sec:decaypattern}

In this section, we determine the dominant collider signatures in the type-I~2HDM+$a$ model with a non-degenerate BSM Higgs sector. Given that in the 2HDM+$a$ model, mono-$X$ signals such as $h+E_{T, \rm miss}$, $Z+E_{T, \rm miss}$, and $tW+E_{T, \rm miss}$ are known to be resonantly enhanced~\cite{No:2015xqa,Bauer:2017ota,Pani:2017qyd}, we will also direct our focus towards processes of this kind. Specifically, we will identify the dominant collider signature as the final state with the highest inclusive signal strength, assuming resonant production of a BSM Higgs boson followed by its decay. Since in the NWA, production and decay processes factorise, identifying the largest inclusive signal strength essentially involves determining the maximal branching ratio of a given BSM Higgs boson into a specific final state. We start our discussion by presenting formulae for the partial decay widths of the $A$, $H$, $H^\pm$, and $a$, which are essential for computing the branching ratios of the BSM spin-$0$ states.

\subsection[Decays of pseudoscalar $A$]{Decays of pseudoscalar $\bm{A}$}
\label{sec:heavierpseudoscalar}

Assuming that all possible decay channels are kinematically open and considering the alignment limit, the non-zero partial widths of the tree-level decays of the pseudoscalar $A$ in the type-I~2HDM+$a$ model are given by:
\beq \label{eq:GammaAX}
\begin{split}
\Gamma \left ( A \to \chi \bar \chi \right ) & = \frac{y_\chi^2}{8\pi} \hspace{0.5mm} m_A \hspace{0.5mm} \beta (m_A, m_\chi) \hspace{0.5mm} \sin^2 \theta \,, \\[2mm]
\Gamma \left ( A \to f \bar f \right ) & = \frac{N_c^f \cot^2 \beta}{8\pi} \hspace{0.1mm} \frac{m_f^2}{v^2} \hspace{0.5mm} m_A \hspace{0.5mm} \beta (m_A, m_f) \hspace{0.5mm} \cos^2 \theta \,, \\[2mm]
\Gamma \left ( A \to Z H \right ) & = \frac{1}{16\pi} \hspace{0.5mm} \frac{\lambda^{3/2} (m_A, m_Z, m_H)}{m_A^3 \hspace{0.25mm} v^2} \hspace{0.5mm} \cos^2 \theta \,, \\[2mm]
\Gamma \left ( A \to W^\mp H^\pm \right ) & = \frac{1}{8 \pi} \hspace{0.5mm} \frac{\lambda^{3/2} (m_A, m_W, m_{H^\pm})}{m_A^3 \hspace{0.25mm} v^2} \hspace{0.5mm} \cos^2 \theta \,, \\[2mm]
\Gamma \left ( A \to h a \right ) & = \frac{1}{16 \pi} \frac{\lambda^{1/2} (m_A, m_h, m_a)}{m_A} \, g_{Aha}^2 \,, \\[2mm]
\Gamma \left ( A \to H a \right ) & = \frac{1}{16 \pi} \frac{\lambda^{1/2} (m_A, m_H, m_a)}{m_A} \, g_{AHa}^2 \,.
\end{split}
\eeq
Here $N_c^f = 3 \, (1)$ denotes the relevant colour factor for quarks (leptons) and we have defined the following kinematic functions
\beq \label{eq:betalambda}
\beta (m_1, m_2) = \sqrt{1 - \frac{4 m_2^2}{m_1^2}} \,, \qquad \lambda (m_1, m_2, m_3) = \left ( m_1^2 - m_2^2 - m_3^2 \right )^2 - 4 \hspace{0.25mm} m_2^2 \hspace{0.5mm} m_3^2 \,.
\eeq
The trilinear couplings appearing in~(\ref{eq:GammaAX}) take the form
\beq \label{eq:gAcouplings}
\begin{split}
g_{Aha} & = \frac{1}{m_A \hspace{0.25mm} v} \, \Big [ \hspace{0.5mm} m_h^2 - 2 m_H^2 - m_A^2 + 4 m_{H^\pm}^2 - m_a^2 - 2 \lambda_3 v^2 \\[1mm] & \hspace{1.6cm} + 2 \left ( \lambda_{P1} \cos^2 \beta + \lambda_{P2} \sin^2 \beta \right ) v^2 \hspace{0.5mm} \Big ] \sin \theta \cos \theta \,, \\[2mm]
g_{AHa} & = \frac{1}{m_A \hspace{0.25mm} v} \, \Big [ \hspace{0.5mm} 2 \cot \left (2\beta \right ) \left ( m_h^2 - 2 m_H^2 + 2 m_{H^\pm}^2 - \lambda_3 v^2 \right ) \\[1mm] & \hspace{1.6cm} - \sin \left ( 2 \beta \right ) \left ( \lambda_{P1} - \lambda_{P2} \right ) v^2 \hspace{0.5mm} \Big ] \sin \theta \cos \theta \,.
\end{split}
\eeq
Notice that the result for the $A \to W^\mp H^\pm$ partial decay rate includes both charge combinations and that in the alignment limit the decay $A \to Zh$ is not possible. At the one-loop level, the pseudoscalar $A$ can decay to gluons and photons, but the associated branching ratios are highly suppressed, particularly for the $\tan\beta$ values considered in the benchmarks below, and therefore can be ignored for all practical purposes.

\subsection[Decays of scalar $H$]{Decays of scalar $\bm{H}$}
\label{sec:heavierscalar}

In the case of the scalar $H$ the relevant formulae corresponding to~(\ref{eq:GammaAX}) are 
\beq \label{eq:GammaHX}
\begin{split}
\Gamma \left ( H \to f \bar f \right ) & = \frac{N_c^f \cot^2 \beta}{8\pi} \hspace{0.1mm} \frac{m_f^2}{v^2} \hspace{0.5mm} m_H \hspace{0.5mm} \beta^3 (m_H, m_f) \,, \\[2mm]
\Gamma \left ( H \to Z A \right ) & = \frac{1}{16 \pi} \frac{\lambda^{3/2} (m_H, m_Z, m_A)}{m_H^3 \hspace{0.25mm} v^2} \, \cos^2 \theta \,, \\[2mm]
\Gamma \left ( H \to Z a \right ) & = \frac{1}{16 \pi} \frac{\lambda^{3/2} (m_H, m_Z, m_a)}{m_H^3 \hspace{0.25mm} v^2} \, \sin^2 \theta \,, \\[2mm]
\Gamma \left ( H \to W^\mp H^\pm \right ) & = \frac{1}{8 \pi} \frac{\lambda^{3/2} (m_H, m_W, m_H^\pm)}{m_H^3 \hspace{0.25mm} v^2} \,, \\[2mm]
\Gamma \left ( H \to A A \right ) & = \frac{1}{32 \pi} \, g_{HAA}^2 \, m_H \hspace{0.25mm} \beta ( m_H, m_A ) \,, \\[2mm]
\Gamma \left ( H \to A a \right ) & = \frac{1}{16 \pi} \frac{\lambda^{1/2} (m_H, m_A, m_a)}{m_H} \, g_{HAa}^2 \,, \\[2mm]
\Gamma \left ( H \to a a \right ) & = \frac{1}{32 \pi} \, g_{Haa}^2 \, m_H \hspace{0.25mm} \beta ( m_H, m_a ) \,, \\[2mm]
\Gamma \left ( H \to H^+ H^- \right ) & = \frac{1}{16 \pi} \, g_{HH^+H^-}^2 \hspace{0.25mm} m_H \hspace{0.25mm} \beta ( m_H, m_{H^\pm} ) \,, 
\end{split}
\eeq
where 
\beq \label{eq:gHcouplings}
\begin{split}
g_{HAA} & = \frac{1}{m_H \hspace{0.25mm} v} \, \Big [ \hspace{0.5mm} 2 \cot \left ( 2 \beta \right) \left ( m_h^2 - 2 m_H^2 + 2 m_{H^\pm}^2 - \lambda_3 v^2 \right ) \cos^2 \theta \\[1mm] & \hspace{1.6cm} + \sin \left ( 2 \beta \right ) \left (\lambda_{P1}-\lambda_{P2} \right ) v^2 \sin^2 \theta \hspace{0.5mm} \hspace{0.5mm} \Big ] \,, \\[2mm]
g_{HAa} & = \frac{1}{m_H \hspace{0.25mm} v} \, \Big [ \hspace{0.5mm} 2 \cot \left (2\beta \right ) \left ( m_h^2 - 2 m_H^2 + 2 m_{H^\pm}^2 - \lambda_3 v^2 \right ) \\[1mm] & \hspace{1.6cm} - \sin \left ( 2 \beta \right ) \left ( \lambda_{P1} - \lambda_{P2} \right ) v^2 \hspace{0.5mm} \Big ] \sin \theta \cos \theta \,, \\[2mm]
g_{Haa} & = \frac{1}{m_H \hspace{0.25mm} v} \, \Big [ \hspace{0.5mm} 2 \cot \left ( 2 \beta \right) \left ( m_h^2 - 2 m_H^2 + 2 m_{H^\pm}^2 - \lambda_3 v^2 \right ) \sin^2 \theta \\[1mm] & \hspace{1.6cm} + \sin \left ( 2 \beta \right ) \left (\lambda_{P1}-\lambda_{P2} \right ) v^2 \cos^2 \theta \hspace{0.5mm} \hspace{0.5mm} \Big ] \,, \\[2mm]
g_{HH^+H^-} & = \frac{1}{m_H \hspace{0.25mm} v} \, \Big [ \hspace{0.5mm} 2 \cot \left ( 2 \beta \right) \left ( m_h^2 - 2 m_H^2 + 2 m_{H^\pm}^2 - \lambda_3 v^2 \right ) \hspace{0.0mm} \Big ] \,.
\end{split}
\eeq
The result for the $H \to W^\mp H^\pm$ partial decay rate includes both charge combinations, while the analytic expressions for the functions $\beta (m_1, m_2)$ and $\lambda (m_1, m_2, m_3)$ can be found in~(\ref{eq:betalambda}). Notice that the decays $H \to ZZ$, $H \to W^+ W^-$, and $H \to hh$ are forbidden in the alignment limit. Like in the case of the pseudoscalar~$A$, loop-induced decays of the scalar~$H$ to gluons and photons are strongly suppressed and can hence be neglected in practical applications. 

\subsection[Decays of charged Higgs $H^+$]{Decays of charged Higgs $\bm{H^+}$}
\label{sec:chargedscalar}

The non-vanishing partial decays widths of the charged Higgs boson in the aligned 2HDM+$a$ model of type~I read 
\beq \label{eq:GammaHpX}
\begin{split}
\Gamma \left ( H^+ \to t \bar b \right ) & = \frac{N_c^t \hspace{0.75mm} |V_{tb}|^2 \cot^2 \beta}{8\pi} \hspace{0.25mm} \frac{m_t^2}{v^2} \hspace{0.5mm} m_{H^\pm} \left ( 1 - \frac{m_t^2}{m_{H^\pm}^2} \right )^2 \,, \\[2mm]
\Gamma \left ( H^+ \to H W^+ \right ) & = \frac{1}{16\pi} \hspace{0.25mm} \frac{\lambda^{3/2} (m_{H^\pm}, m_H, m_W)}{m_{{H^\pm}}^3 \hspace{0.25mm} v^2} \,, \\[2mm]
\Gamma \left ( H^+ \to A W^+ \right ) & = \frac{1}{16\pi} \hspace{0.25mm} \frac{\lambda^{3/2} (m_{H^\pm}, m_A, m_W)}{m_{{H^\pm}}^3 \hspace{0.25mm} v^2} \, \cos^2 \theta \,, \\[2mm]
\Gamma \left ( H^+ \to a W^+ \right ) & = \frac{1}{16\pi} \hspace{0.25mm} \frac{\lambda^{3/2} (m_{H^\pm}, m_a, m_W)}{m_{{H^\pm}}^3 \hspace{0.25mm} v^2} \, \sin^2 \theta \,, 
\end{split}
\eeq
where in the case of the $H^+ \to t \bar b$ decay we have neglected terms of ${\cal O} (m_b^2/M_{H^\pm}^2)$ in the expression for the partial decay width. The results for $\Gamma \left ( H^+ \to c \bar s \right )$ and $\Gamma \left ( H^+ \to \tau^+ \nu \right )$ follow from the expression for $\Gamma \left ( H^+ \to t \bar b \right )$ after obvious replacements. Notice that the $H^+ \to h W^+$ decay is absent in the case of alignment. 

\subsection[Decays of pseudoscalar $a$]{Decays of pseudoscalar $\bm{a}$}
\label{sec:lightpseudoscalar}

Assuming that the decays to final states involving a $A$, $H$ or $H^\pm$ are kinematically inaccessible, the non-zero partial decay widths of the pseudoscalar $a$ are given in the aligned type-I~2HDM+$a$ model by
\beq \label{eq:GammaaX}
\begin{split}
\Gamma \left ( a \to \chi \bar \chi \right ) & = \frac{y_\chi^2}{8\pi} \hspace{0.5mm} m_a \hspace{0.5mm} \beta (m_a, m_\chi) \hspace{0.5mm} \cos^2 \theta \,, \\[2mm]
\Gamma \left ( a \to f \bar f \right ) & = \frac{N_c^f \cot^2 \beta}{8\pi} \hspace{0.1mm} \frac{m_f^2}{v^2} \hspace{0.5mm} m_a \hspace{0.5mm} \beta (m_a, m_f) \hspace{0.5mm} \sin^2 \theta \,.
\end{split}
\eeq
It is important to realise that the decay rate of $a \to Zh$ vanishes identically in the limit $\cos \left ( \beta - \alpha \right ) = 0$. At the one-loop level the pseudoscalar $a$ can also decay to gauge bosons. The largest partial decay width is the one to gluon pairs. The relevant formulae can be found in~Section~4.1 of~\cite{Bauer:2017ota}. They will not be reproduced here because, similar to the case of $A$ and $H$, loop decays of the pseudoscalar $a$ are suppressed for the sizeable $\tan\beta$ values considered below.

\subsection{Dominant collider signatures}
\label{sec:signatures}

\begin{figure}[t!]
\begin{center}
\includegraphics[width=0.99\textwidth]{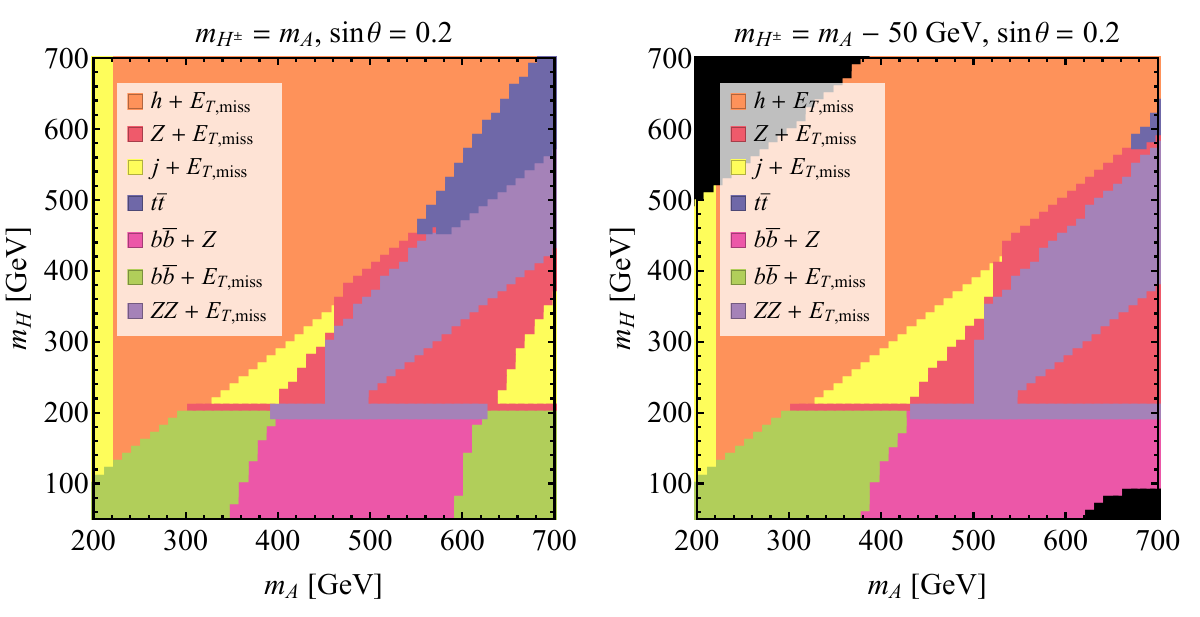}

\includegraphics[width=0.99\textwidth]{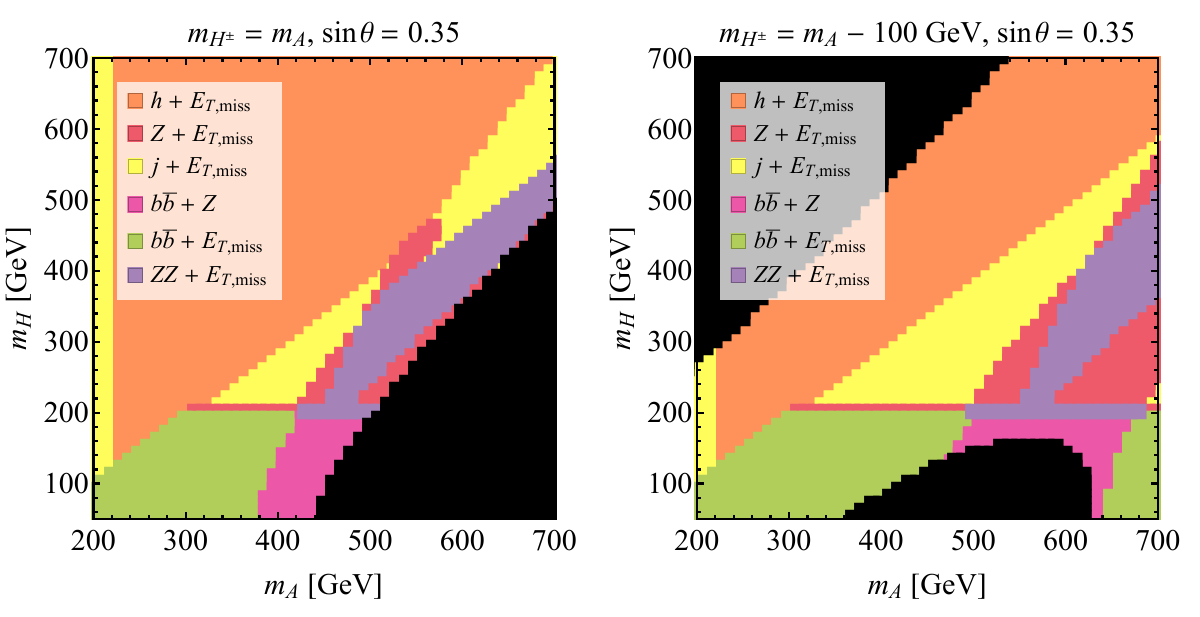}
\vspace{0mm}
\caption{\label{fig:muA} Dominant final-state configurations due to resonant ggF production of the pseudoscalar~$A$ in the $m_A\hspace{0.125mm}$--$\hspace{0.5mm}m_H$ plane. Four different choices of $m_{H^\pm}$ and $\sin \theta$, as shown in the headline of the panels, are depicted. The 2HDM+$a$ input parameters not indicated are $m_a = 100 \, {\rm GeV}$, $m_\chi = 10 \, {\rm GeV}$, $\cos \left ( \beta - \alpha \right ) = 0$, $\tan \beta = 5$, $y_\chi = 1$, $\lambda_3 = 6$, $\lambda_{P1, P2} = 0$, and $\lambda_{P3} > 0$. The~black regions in the parameter space are incompatible with the constraint~(\ref{eq:rhoconstraint}) on the~$\rho$~parameter. See main text for additional explanations.} 
\end{center}
\end{figure}

Using the formulae given in the four preceding sections we are now able to determine the dominant collider signatures that one expects to observe in the type-I~2HDM+$a$ model with a non-degenerate BSM Higgs sector. For this purpose, we identify the final state~$X$ with the highest inclusive signal strength $\sigma \left ( g g \to \Phi \right ) {\rm Br} \left (\Phi \to X \right )$, assuming resonant production of $\Phi = A, H$ in ggF. Figure~\ref{fig:muA} shows the dominant final-state configurations initiated by resonant production of the pseudoscalar $A$ in the $m_A\hspace{0.125mm}$--$\hspace{0.5mm}m_H$ plane, considering four distinct choices of $m_{H^\pm}$ and $\sin \theta$. The~input parameters not indicated in the headlines of the panels are $m_a = 100 \, {\rm GeV}$, $m_\chi = 10 \, {\rm GeV}$, $\cos \left ( \beta - \alpha \right ) = 0$, $\tan \beta = 5$, $y_\chi = 1$, $\lambda_3 = 6$, and $\lambda_{P1,P2} = 0$. Notice that the choice of $m_a$, $m_\chi$, and $y_{\chi}$ leads to ${\rm Br} \left ( a \to \chi \bar \chi \right ) \simeq 100\%$ meaning that final states containing a pseudoscalar $a$ represent a potentially interesting $E_{T, \rm miss}$ signature. Moreover, the choices $\cos \left ( \beta - \alpha \right ) = 0$ and $\tan \beta = 5$ curb the constraints imposed by Higgs and  flavour physics as well as direct collider searches. In contrast, the choice $\lambda_3 = 6$ is specifically designed to satisfy the tree-level BFB conditions, particularly in scenarios involving large mass splittings within the BSM Higgs sector, while $\lambda_{P1, P2} = 0$ keeps the total decay widths of the BSM spin-$0$ states small. A positive value of $\lambda_{P3}$, which is required by the BFB conditions but does not affect the results shown, is understood. See Section~\ref{sec:constraints} for further details. The~parameter space that is incompatible with the constraint~(\ref{eq:rhoconstraint}) on the $\rho$~parameter is indicated in black in the four panels of the figure. From the plots in~Figure~\ref{fig:muA} it is evident that in the studied parameter scenarios there are seven different signatures that can provide the dominant final state in $A$ decays:

\begin{figure}[t!]
\begin{center}
\includegraphics[width=0.8\textwidth]{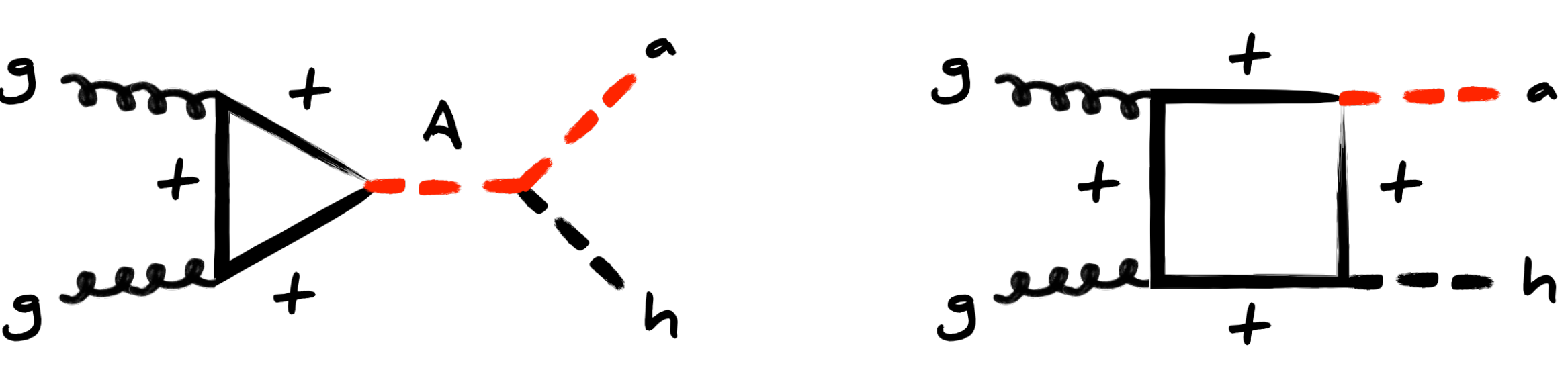}
\vspace{2mm}
\caption{\label{fig:diagrams1} Examples of Feynman diagrams that can give rise to a $h + E_{T, \rm miss}$ signature in the type-I~2HDM+$a$ model with a non-degenerate BSM Higgs sector. The decays $a \to \chi \bar \chi$ that lead to the $E_{T, \rm miss}$ signal are not shown for simplicity.} 
\end{center}
\end{figure}

\begin{figure}[t!]
\begin{center}
\includegraphics[width=0.9\textwidth]{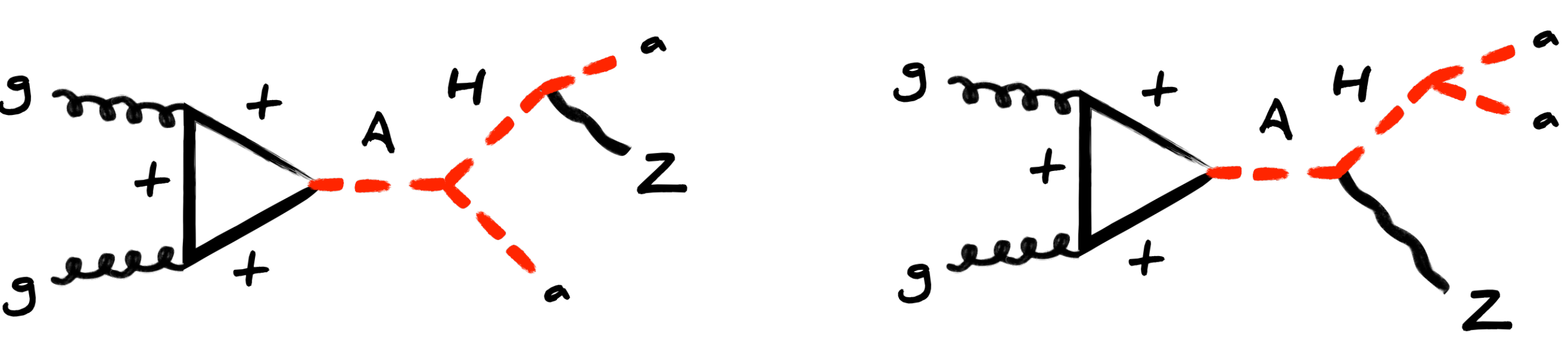}
\vspace{2mm}
\caption{\label{fig:diagrams2} As Figure~\ref{fig:diagrams1} but for $Z + E_{T, \rm miss}$ production. For further explanations see the main text.} 
\end{center}
\end{figure}

\begin{figure}[t!]
\begin{center}
\includegraphics[width=0.825\textwidth]{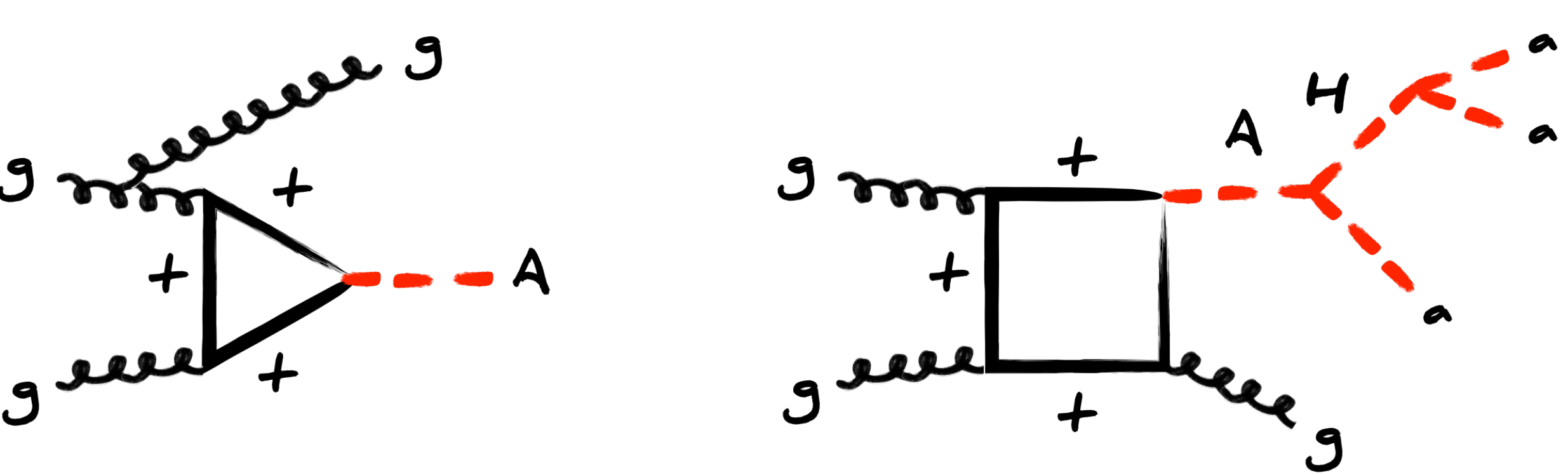}
\vspace{2mm}
\caption{\label{fig:diagrams3} As Figure~\ref{fig:diagrams1} but for $j + E_{T, \rm miss}$ production. Notice that the decays $A \to \chi \bar \chi$ and $a \to \chi \bar \chi$ both give rise to a $E_{T, \rm miss}$ signature. In addition to the left diagram, the process $gg\to a$ is also present, and in fact dominates $j + E_{T, \rm miss}$ production for $m_A>m_a$ and $\sin \theta$ sufficiently large.} 
\end{center}
\end{figure}

\begin{figure}[t!]
\begin{center}
\includegraphics[width=0.9\textwidth]{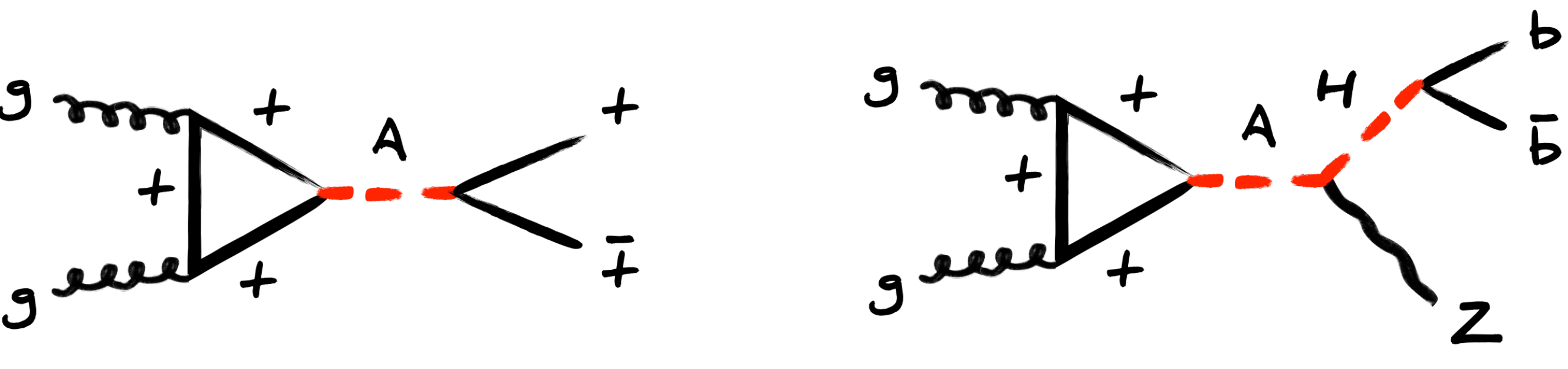}
\vspace{2mm}
\caption{\label{fig:diagrams4} Resonant contributions to $t \bar t$~(left) and $b \bar b + Z$~(right) production in the 2HDM+$a$ model of type~I with a non-degenerate BSM Higgs sector. } 
\end{center}
\end{figure}

\begin{figure}[t!]
\begin{center}
\includegraphics[width=0.925\textwidth]{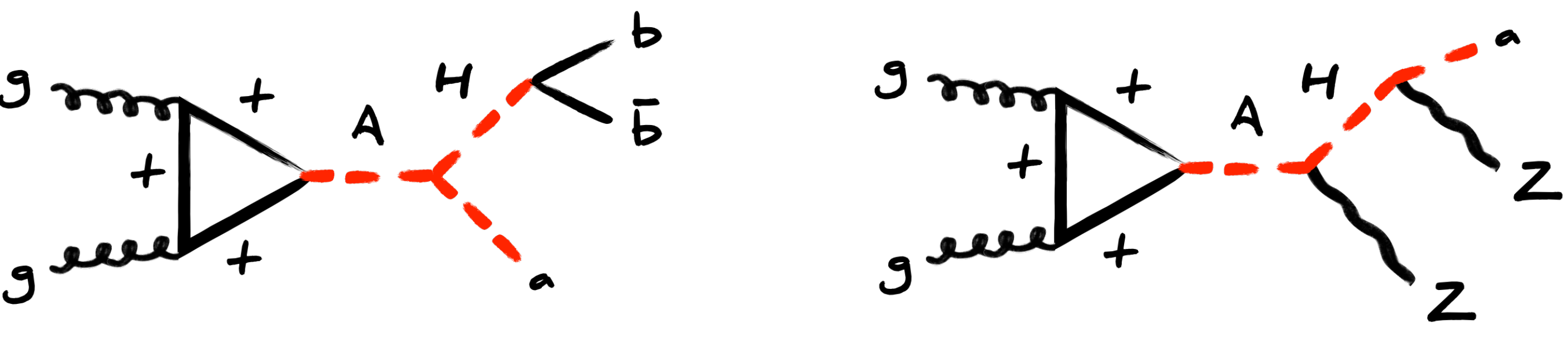}
\vspace{2mm}
\caption{\label{fig:diagrams5} As Figure~\ref{fig:diagrams4} but for $b \bar b + E_{T, \rm miss}$~(left) and $ZZ + E_{T, \rm miss}$~(right) production.} 
\end{center}
\end{figure}

\begin{itemize}

\item[{\color{color1} $\blacksquare$}]{$h + E_{T, \rm miss}$:} Relevant resonant and non-resonant Feynman graphs that can lead to mono-Higgs production in the type-I~2HDM+$a$ model with a non-degenerate BSM Higgs sector are depicted in~Figure~\ref{fig:diagrams1}. This specific signature has been searched for by ATLAS and~CMS, and interpreted in the context of the 2HDM+$a$ model of type~II~in~\cite{CMS:2018zjv,ATLAS:2019wdu,ATLAS:2021jbf,ATLAS:2021shl,CMS:2022sfl,ATLAS:2023ild}. In~Section~\ref{sec:hMETsearch}, we present the constraints on the type-I~2HDM+$a$ model from the $h + E_{T, \rm miss}$ signature in the $h \to b \bar b$ channel using a truth-level analysis. We will see that $h + E_{T, \rm miss}$ searches at the LHC can, indeed, set relevant bounds in parts of the $m_A\hspace{0.125mm}$--$\hspace{0.5mm}m_H$ planes with $m_h + m_a \leq m_A \leq m_H + m_a$, where the resonant $A\to ah$ process is enhanced.

\item[\color{color2} $\blacksquare$]{$Z + E_{T, \rm miss}$:} Example diagrams that contribute to resonant mono-$Z$ production in the type-I~2HDM+$a$ model with a non-degenerate BSM Higgs sector are displayed in~Figure~\ref{fig:diagrams2}. Notice that the shown graphs describe the $gg \to A \to H a \to Z aa$ and $gg \to A \to H Z \to Z aa$ processes, which are not allowed if $A$ and $H$ are mass-degenerate. The shape of the resulting kinematic distributions therefore differs from those discussed in Section~6.1.2 of the whitepaper \cite{LHCDarkMatterWorkingGroup:2018ufk}, which focused on the~2HDM+$a$ model of type~II featuring a mass-degenerate spectrum of $A$, $H$, and $H^\pm$. In that scenario, the $gg \to H \to Za$ channel typically dominates $Z + E_{T, \rm miss}$ production~\cite{No:2015xqa,Bauer:2017ota}. Still using the information on the existing mono-$Z$ searches~\cite{ATLAS:2019wdu,CMS:2020ulv,ATLAS:2021gcn} and applying the methodology described in~Section~8.2~of~\cite{LHCDarkMatterWorkingGroup:2018ufk} one can estimate the expected sensitivity of the $Z + E_{T, \rm miss}$ signature to the relevant parameter space in the $m_A\hspace{0.125mm}$--$\hspace{0.5mm}m_H$ planes of~Figure~\ref{fig:diagrams1}. We~comment on the LHC reach of $Z + E_{T, \rm miss}$ searches within the type-I~2HDM+$a$ model in~Section~\ref{sec:others}.

\item[\color{color3} $\blacksquare$]{$j + E_{T, \rm miss}$:} Graphs that lead to a mono-jet signature in the type-I~2HDM+$a$ model with a non-degenerate BSM Higgs sector are depicted in~Figure~\ref{fig:diagrams3}. One sees that there are two types of $j + E_{T, \rm miss}$ contributions that result from the decay of the pseudoscalar $A$, namely the processes $gg \to A$ and $gg \to A \to H a \to aaa$. Notice that in the 2HDM+$a$ model of type~I there is also always a mono-jet contribution associated to $gg \to a$ production. In fact, in the $m_A\hspace{0.125mm}$--$\hspace{0.5mm}m_H$ planes shown in~Figure~\ref{fig:diagrams1} the latter channel always provides the largest contribution to mono-jet production because of $m_A > m_a$. The regions in parameter space in which the $j + E_{T, \rm miss}$ signature dominates can be targeted by searches for energetic jets and large $E_{T, \rm miss}$ such as~\cite{ATLAS:2021kxv,CMS:2021far}. 

\item[\color{color4} $\blacksquare$]{$t \bar t$:} In the type-I~2HDM+$a$ model with a non-degenerate BSM Higgs sector, $A$ decays can also lead to a $t \bar t$ signal. The relevant diagram is shown on the left-hand side in~Figure~\ref{fig:diagrams4}. Model realisations that predict a sufficiently large $t \bar t$ signal can be tested by studying the invariant mass spectrum $m_{t \bar t}$. Interference effects between the signal process and the SM $t \bar t$ background, however, distort the $m_{t \bar t}$ signal shape from a single peak to a peak-dip structure~\cite{Gaemers:1984sj,Dicus:1994bm,Bernreuther:1997gs,Frederix:2007gi,Hespel:2016qaf,BuarqueFranzosi:2017jrj}. This feature represents a generic obstacle to test 2HDM models and as a result the existing $t \bar t$ resonance searches~\cite{ATLAS:2017snw,CMS:2019pzc,ATLAS-CONF-2024-001} are only able to exclude $\tan \beta \lesssim 3.5$ for $m_{A} = m_H = 400 \, {\rm GeV}$. Since the parameter scenarios studied in~Figure~\ref{fig:muA} all employ $\tan \beta = 5$ and $\sin \theta \ll 1/\sqrt{2}$, searches for resonant $t \bar t$~production do not presently set any constraint on the considered 2HDM+$a$~models of~type~I --- see~\cite{ATLAS-CONF-2024-001} for the ATLAS 2HDM+$a$ interpretations of their latest $t \bar t$ resonance search.

\item[\color{color5} $\blacksquare$]{$b \bar b + Z$:} The diagram on the right-hand side of~Figure~\ref{fig:diagrams4} represents a relevant contribution to $b \bar b + Z$ production in the type-I~2HDM+$a$ model with a non-degenerate BSM Higgs sector. The signal arises from the decay $A \to ZH$ and this channel has been searched for by ATLAS and CMS both in the $\ell^+ \ell^- b \bar b$~\cite{ATLAS:2018oht,CMS:2019ogx,ATLAS:2020gxx} and the $\nu \bar \nu b \bar b$~\cite{ATLAS:2023zkt} final states. A dedicated analysis strategy that targets the $b \bar b + Z$ signal is described in~Section~\ref{sec:ZHsearches}. As we will see, our search strategy can be used to constrain 2HDM+$a$ model realisations of the type studied in~Figure~\ref{fig:muA} with relatively light $H$ of $m_H \lesssim 200 \, {\rm GeV}$. For $m_H \gtrsim 200 \, {\rm GeV}$, the sensitivity of the $b \bar b + Z$ signature is reduced, either due to the $A\to ZH$ decay becoming subdominant or because the $H$ preferentially decays via $H \to Z a$ or $H \to aa$.

\item[\color{color6} $\blacksquare$]{$b \bar b + E_{T, \rm miss}$:} A possible contribution to $b \bar b + E_{T, \rm miss}$ production in the context of the type-I~2HDM+$a$ model is displayed on the left in~Figure~\ref{fig:diagrams5}. Notice that the $b \bar b$ pair in this case does not arise from the decay of the $125 \, {\rm GeV}$ Higgs boson but the scalar~$H$ that appears in the decay chain $gg \to A \to H a \to b \bar b a$. In consequence, the kinematic distributions of the $b \bar b + E_{T, \rm miss}$ signal, like for instance, the invariant mass of the~$b \bar b$~system are different from that of the usual 2HDM+$a$ mono-Higgs signature in the $b \bar b$~channel resulting from $gg \to A \to h a \to b \bar b a$. It is noteworthy that in addition to $A\to Ha$, the $A\to ZH$ process, followed by $Z \to \nu \bar \nu$, yields the same signature, with the latter dominating at low $m_A$. To study the sensitivity of Run 2 data to the $b \bar b + E_{T, \rm miss}$ signature, we develop a truth-level analysis, following~\cite{ATLAS:2023zkt}. Our search is described in detail in~Section~\ref{sec:hMETsearch}, and we find that it allows to test model realisations with $m_H \lesssim 200 \, {\rm GeV}$ for a wide range of $m_A$ values. We add that while the region $m_H \geq 125 \, {\rm GeV}$ has been investigated in~\cite{ATLAS:2023zkt}, the mass range $m_H < 125 \, {\rm GeV}$ remains unexplored. In principle, a light $b \bar b$ resonance could also emerge in models with two mediators~\cite{Duerr:2016tmh}, which have been examined for instance in~\cite{ATLAS:2019ivx,ATLAS-CONF-2024-004}. However, the production mechanism and, consequently, the signal kinematics are expected to be significantly different from the case studied here.

\item[\color{color7} $\blacksquare$]{$Z Z + E_{T, \rm miss}$:} The graph on the right in~Figure~\ref{fig:diagrams5} represents a resonant contribution to $Z Z + E_{T, \rm miss}$ production in the type-I~2HDM+$a$ model. This signal arises from the process $gg \to A \to ZH \to ZZa$ in a patch of parameter space with $m_A \gtrsim 500 \, {\rm GeV}$ and $m_H \gtrsim 200 \, {\rm GeV}$. The $Z Z + E_{T, \rm miss}$ signature can be constrained by LHC searches such as the analysis~\cite{ATLAS:2021wob} that looks for four-lepton~($4\ell$) events with $E_{T, \rm miss}$ where the leptons originate from $Z$-boson candidates --- see very recently also~\cite{ATLAS:2024bzr}. Instead of relying on~\cite{ATLAS:2021wob,ATLAS:2024bzr}, which utilise the invariant mass of the $4\ell$ system, we develop our own truth-level analysis in~Section~\ref{sec:ZZMETsearches}. Our analysis employs the transverse mass of the leptons and $E_{T, \rm miss}$, and notably, it yields superior constraints on $ZZ + E_{T, \rm miss}$ production compared to simple reinterpretations of~\cite{ATLAS:2021wob,ATLAS:2024bzr}. Through our analysis, we find that for the four benchmarks depicted in~Figure~\ref{fig:muA}, searches for a $ZZ + E_{T, \rm miss}$ signal using all data available after LHC~Run~3 can exhibit sensitivity to the parameter space with $m_A \simeq 500 \, {\rm GeV}$ and $m_H \simeq 300 \, {\rm GeV}$.

\end{itemize}

\begin{figure}[t!]
\begin{center}
\includegraphics[width=0.99\textwidth]{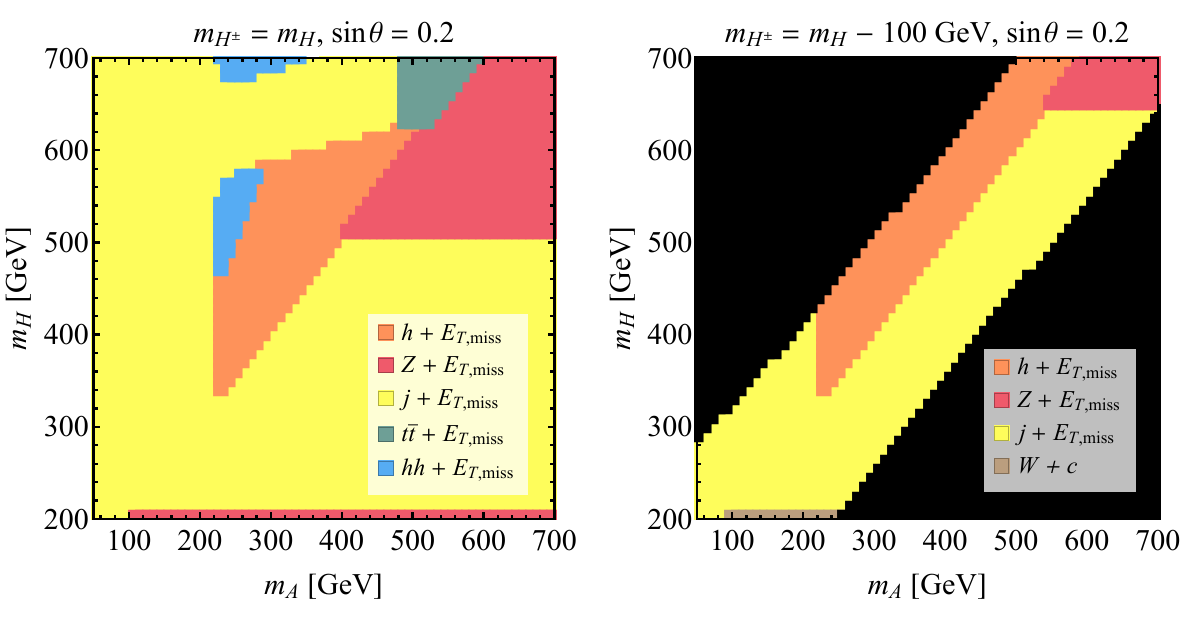}

\includegraphics[width=0.99\textwidth]{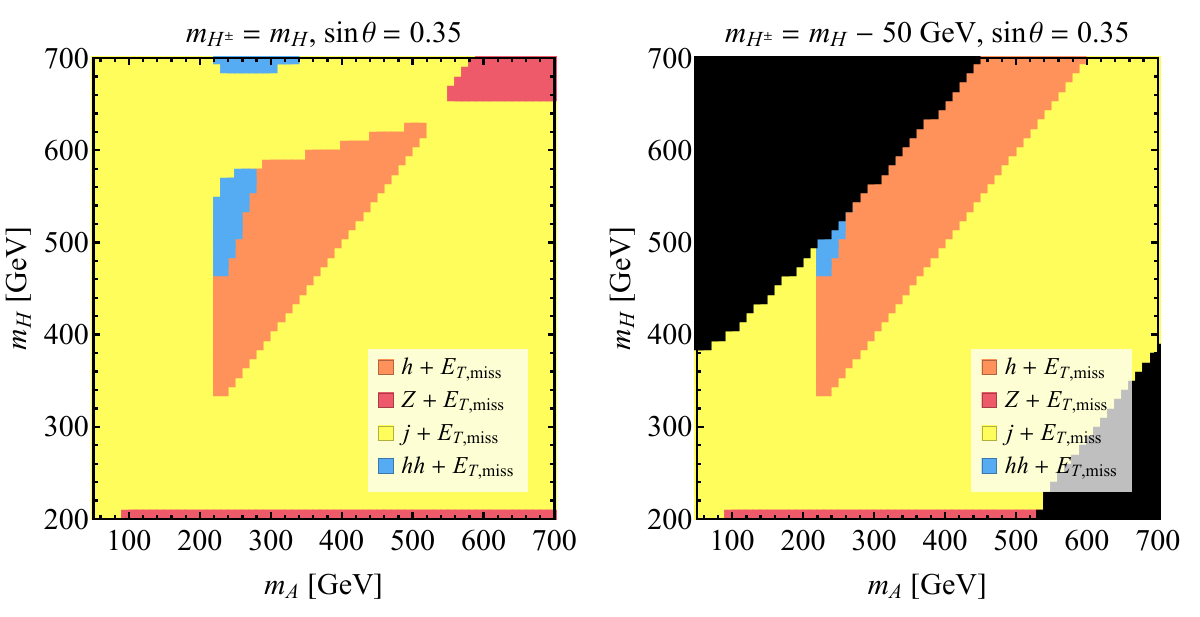}
\vspace{0mm}
\caption{\label{fig:muH} As Figure~\ref{fig:muA} but for resonant ggF production of the scalar~$H$. Detailed explanations can be found in the main text.} 
\end{center}
\end{figure}

Figure~\ref{fig:muH} shows the dominant final-state configurations resulting from the resonant production of the scalar $H$ in the $m_A\hspace{0.125mm}$--$\hspace{0.5mm}m_H$ plane for four different choices of $\sin \theta$ and~$m_{H^\pm}$. The~2HDM+$a$ input parameters not indicated in the headings are identical to those used to obtain~Figure~\ref{fig:muA}. We observe that the process $gg \to H$ leads to five different final states that can be relevant for collider phenomenology:

\begin{figure}[t!]
\begin{center}
\includegraphics[width=0.9\textwidth]{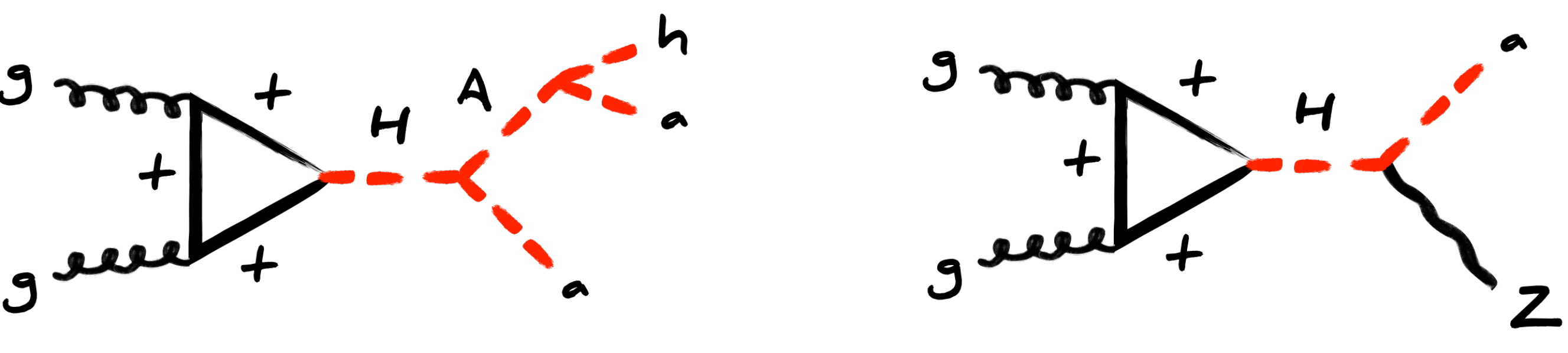}
\vspace{2mm}
\caption{\label{fig:diagrams6} Contribution to $h + E_{T, \rm miss}$~(left) and $Z + E_{T, \rm miss}$~(right) production in the type-I~2HDM+a model with a non-degenerate BSM Higgs sector that arise from the decay of the scalar $H$. The $E_{T, \rm miss}$ signal stems from the decay $a \to \chi \bar \chi$. See main text for further details.} 
\end{center}
\end{figure}

\begin{figure}[t!]
\begin{center}
\includegraphics[width=0.9\textwidth]{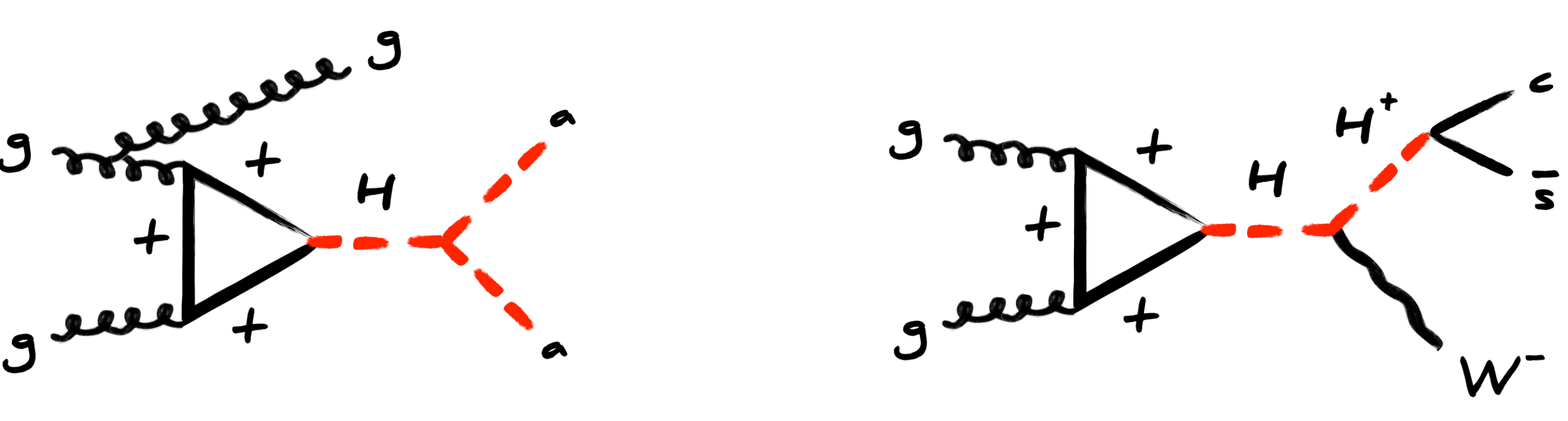}
\vspace{2mm}
\caption{\label{fig:diagrams7} As Figure~\ref{fig:diagrams6} but for $j+ E_{T, \rm miss}$~(left) and $W + c$~(right) production.}
\end{center}
\end{figure}

\begin{figure}[t!]
\begin{center}
\includegraphics[width=0.9\textwidth]{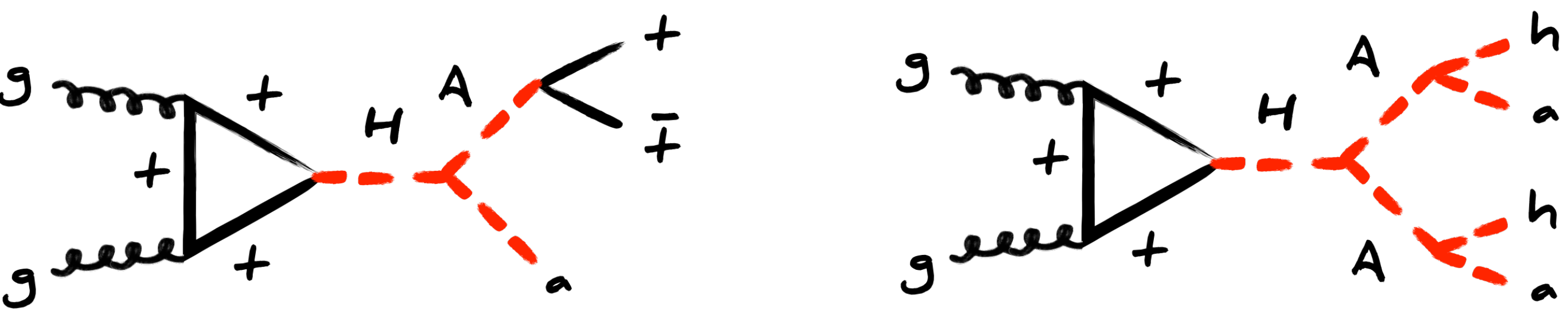}
\vspace{2mm}
\caption{\label{fig:diagrams8} As Figure~\ref{fig:diagrams6} but for $t \bar t + E_{T, \rm miss}$~(left) and $hh + E_{T, \rm miss}$~(right) production.}
\end{center}
\end{figure}

\begin{itemize}

\item[{\color{color1} $\blacksquare$}]{$h + E_{T, \rm miss}$:} A contribution to mono-Higgs production in the type-I~2HDM+$a$ model with a non-degenerate BSM Higgs sector is given on the left in~Figure~\ref{fig:diagrams6}. Notice~that unlike in~Figure~\ref{fig:diagrams1} the $h + E_{T, \rm miss}$ signature arises from $gg \to H \to A a \to haa$. As~before, one can use the analysis strategy described in~Section~\ref{sec:hMETsearch} to constrain the wedge in the $m_A\hspace{0.125mm}$--$\hspace{0.5mm}m_H$ planes centred around $m_A \simeq 300 \, {\rm GeV}$ and $m_H \simeq 500 \, {\rm GeV}$. We~will comment on the impact of this type of $h + E_{T, \rm miss}$ contributions in~Section~\ref{sec:constraintssummary} and Appendix~\ref{app:h4chi}.

\item[\color{color2} $\blacksquare$]{$Z + E_{T, \rm miss}$:} Besides the contributions from the diagrams in~Figure~\ref{fig:diagrams4}, the $Z + E_{T, \rm miss}$ process also receives contributions from the production of a $H$ boson, which further decays into a $Z$ boson and an $A$ or $a$ boson, as depicted on the right-hand side in~Figure~\ref{fig:diagrams6}. The~results of~\cite{ATLAS:2019wdu,CMS:2020ulv,ATLAS:2021gcn}, which target $gg\to H\to Za$ production, can therefore be utilised to constrain the parameter space in the $m_A\hspace{0.125mm}$--$\hspace{0.5mm}m_H$ planes with $m_{A} \simeq m_H  \simeq 650 \, {\rm GeV}$.

\item[\color{color3} $\blacksquare$]{$j + E_{T, \rm miss}$:} Mono-jet production in the type-I~2HDM+$a$ model that involves the decay of a $H$ is displayed on the left in~Figure~\ref{fig:diagrams7}. Apart from the $gg \to H \to aa$ process also the $gg \to H \to AA$ and $gg \to H \to Aa$ transitions can be relevant. Standard searches for energetic jets and large $E_{T, \rm miss}$ like~\cite{ATLAS:2021kxv,CMS:2021far} can be used to target the regions in parameter space in~Figure~\ref{fig:muH} in which the $j + E_{T, \rm miss}$ signal is important. 

\item[\color{color8} $\blacksquare$]{$W + c$:} The right diagram in~Figure~\ref{fig:diagrams7} represents a contribution to $W + c$ production in the type-I~2HDM+$a$ model with non-degenerate BSM Higgs bosons. The underlying reaction that induces the signal is $gg \to H \to W^- H^+ \to W^- c \bar s$. Notice that the $W + c$ signature is only viable and relevant in a very narrow parameter range with $100 \, {\rm GeV} \lesssim m_A \lesssim 250 \, {\rm GeV}$, $m_H \simeq 200 \, {\rm GeV}$, and $m_{H^\pm} \simeq 100 \, {\rm GeV}$. As we will see in~Section~\ref{sec:chargedHiggssearches}, the LHC searches for the $H^+ \to c \bar s$ or $H^+ \to \tau^+ \nu$ decays~\cite{CMS:2015yvc,CMS:2020osd,ATLAS:2018gfm,CMS:2019bfg} impose $m_H^\pm > 160 \, {\rm GeV}$ for $\tan \beta = 5$ and thus exclude the predicted $W + c$ signature. 

\item[\color{color9} $\blacksquare$]{$t \bar t + E_{T, \rm miss}$:} A graph that can contribute to $t \bar t + E_{T, \rm miss}$ production in the 2HDM+$a$ model of type~I is depicted on the left-hand side of~Figure~\ref{fig:diagrams8}. Notice that besides the one-loop contribution $gg \to H \to A a \to t \bar t a$ that involves the decay of a $H$ there are also the tree-level contributions $gg \to t \bar t A$ and $gg \to t \bar t a$ that lead to a $t \bar t + E_{T, \rm miss}$ signal in the type-I~2HDM+$a$ model. In the type-II 2HDM+$a$ model investigated in~\cite{LHCDarkMatterWorkingGroup:2018ufk,Bauer:2017ota}, the $t\bar{t} + E_{T, \rm miss}$ signature only emerges in the non-resonant $gg\to t\bar{t} A \to t\bar{t} \chi \bar{\chi}$ and $gg\to t\bar{t} a \to t\bar{t} \chi \bar{\chi}$ processes. This is due to the assumption of a degenerate BSM Higgs boson spectrum, resulting in smaller sensitivity compared to resonant signatures. Conversely, in type-I 2HDM+$a$ models, resonant $t \bar{t} + E_{T, \rm miss}$ production could offer relevant constraints for specific benchmarks in the high $m_H$ region. Variants of searches for two top quarks and $E_{T, \rm miss}$ in the final state \cite{ATLAS:2021hza,CMS:2021eha,ATLAS:2022ygn,ATLAS:2024rcx} can target the $t \bar{t} + E_{T, \rm miss}$ signal depicted in~Figure \ref{fig:muH}.

\item[\color{color10} $\blacksquare$]{$hh + E_{T, \rm miss}$:} The diagram on the right in~Figure~\ref{fig:diagrams8} gives rise to a $hh + E_{T, \rm miss}$ signal in the type-I~2HDM+$a$ model with non-degenerate BSM Higgs sector. As one can see the~$hh + E_{T, \rm miss}$ final state arises in our case from the $gg \to H \to A A \to hhaa$ process. A double-Higgs plus $E_{T, \rm miss}$ signature of this type has been investigated in a simplified model in~\cite{Blanke:2019hpe,Flores:2019hcf}. Experimental searches for $hh + E_{T, \rm miss}$ production at the LHC have been performed in the $h h \to b \bar b b \bar b$~\cite{CMS:2017nin,ATLAS:2018tti,CMS:2022vpy} and $h h \to b \bar b \gamma \gamma$~\cite{ATLAS:2023rie} channel. Reinterpreting the existing searches in terms of the type-I~2HDM+$a$ model is, however, not straightforward given that the ATLAS and CMS analyses are optimised for massless DM particles. In~Section~\ref{sec:others}, we briefly discuss the potential reach of $hh + E_{T, \rm miss}$ searches at the LHC within the type-I~2HDM+$a$ model. 
 
\end{itemize}

The discussion above illustrates that in the 2HDM+$a$ model of type~I with a non-degenerate BSM Higgs sector, intriguing additional signatures can emerge. Examples include $b \bar b + Z$, $b \bar b + E_{T, \rm miss}$, $Z Z + E_{T, \rm miss}$, and $hh + E_{T, \rm miss}$, which have not been investigated within the context of the 2HDM+$a$ model until now. The ensuing discussion will focus on these novel signals.

\section{Type-I~2HDM+$\bm{a}$ benchmark models}
\label{sec:benchmarks}

The discussion of the experimental and theoretical constraints presented in~Section~\ref{sec:constraints} together with our explicit study of the decay pattern of the BSM spin-$0$ states~in~Section~\ref{sec:decaypattern} suggest certain benchmarks for the parameters given on the right in~(\ref{eq:2HDMainput}). This section describes how the parameter space of the type-I~2HDM+$a$ model can be effectively explored and constrained through two-dimensional (2D) scans. The benchmark scenarios proposed in this article are not exhaustive scans of the entire type-I~2HDM+$a$ parameter space but are intended to spotlight experimental signatures that remain unexplored at the LHC, to showcase their complementarity and to ensure a consistent comparison of results across different analyses. The parameter choices that are common to the benchmarks studied in the following are: 
\beq \label{eq:generalparameter}
\begin{split}
& m_a = 100 \, {\rm GeV} \,, \qquad m_\chi = 10 \, {\rm GeV} \,, \qquad \cos \left ( \beta - \alpha \right ) = 0 \,, \qquad \tan \beta = 5 \,, \\[2mm]
& \hspace{2cm} y_\chi = 1 \,, \qquad \lambda_3 = 6 \,, \qquad \lambda_{P1} = 0 \,, \qquad \lambda_{P2} = 0 \,.
\end{split}
\eeq
In addition, we assume $\lambda_{P3} > 0$ throughout, as required by the BFB conditions discussed in~Section~\ref{sec:theoryconstraints}. Since $\lambda_{P3}$ affects neither the masses, mixings, trilinear couplings nor decay widths of the spin-$0$ states at tree level, its precise value has no impact on any of the results presented in~Sections~\ref{sec:decaypattern} and~\ref{sec:results}.

The main 2D parameter grid proposed to explore the type-I~2HDM+$a$ model with LHC data spans the combination of the masses $m_A$ and $m_H$. This choice is also employed in many 2HDM interpretations of ATLAS and CMS such as~\cite{ATLAS:2018oht,CMS:2019ogx,ATLAS:2020gxx,ATLAS:2023zkt,ATLAS:2024bzr,CMS:2016xnc}. Example scans in the suggested mass-mass plane have already been given in~Figure~\ref{fig:muA} and Figure~\ref{fig:muH}, where different choices for $m_{H^\pm}$ and $\sin \theta$ have been used in each panel. Our sensitivity studies performed in the next section will rely on the same 2D scans and benchmark parameter choices. Notice that one could also choose to work in the mass plane spanned by $m_A$ and $m_A - m_{H^\pm}$ or $m_H$ and $m_H - m_{H^\pm}$. These representations prove advantageous when one is primarily focused on the constraints arising from EW precision measurements. This is because parameter variations within these 2D planes effectively explore the $\rho$ parameter. However, from a LHC perspective 2D scans in the $m_A\hspace{0.125mm}$--$\hspace{0.5mm}m_H$~plane seem more natural given that the most interesting collider signatures stem from resonant~$A$~and~$H$ production followed by the decays of the neutral BSM Higgs bosons. 

In all the 2D scans conducted in this study, the values of $m_a$, $m_\chi$, and $\tan \beta$ will be held fixed to those specified in~(\ref{eq:generalparameter}). However, allowing the latter parameters to vary may also be of interest. For~example, when considering somewhat heavier pseudoscalars $a$ with $m_a = 300 \, {\rm GeV}$, specific parameter choices exist that can lead to a potentially observable $gg \to A \to W^\mp H^\pm \to W^\mp tb$ signal. The relevance of resonant $gg \to A \to W^\mp H^\pm$ production has been pointed out in the context of the pure 2HDM model of type I with a light scalar $H$ already in~\cite{Haisch:2017gql}. Similar to the $b \bar b + Z$, $b \bar b + E_{T, \rm miss}$, $Z Z + E_{T, \rm miss}$, and $hh + E_{T, \rm miss}$ final states, the latter process represents a novel LHC signature for which, to the best of our knowledge, no dedicated searches by ATLAS or CMS currently exist. We believe that this deserves further study. In order to establish a connection with DM phenomenology, such as relic density calculations, which heavily rely on the DM mass, it would also be instructive to conduct 2D scans in the $m_a\hspace{0.125mm}$--$\hspace{0.5mm}m_\chi$ plane while keeping the values of $m_A$, $m_H$, $m_{H^\pm}$, and $\tan \beta$ fixed. We provide a short overview of the behaviour of the~relic density in~Section~\ref{sec:relicdensity}, however a complete description of the DM phenomenology of the type-I~2HDM+$a$ model in the cosmological context exceeds the scope of this article.

\section{LHC sensitivity studies}
\label{sec:results}

In this section, we provide sensitivity assessments for several novel collider signals within the context of the type-I~2HDM+$a$ model, featuring a non-degenerate BSM Higgs sector. Specifically, we examine the constraints that arise from existing LHC searches for light charged Higgs bosons and the bounds that derive from searches for the $h + E_{T, \rm miss}$, $b \bar b + Z$, $b \bar b + E_{T, \rm miss}$, and $Z Z + E_{T, \rm miss}$ channels. In the latter four cases, we design our own analysis strategies and use them to obtain sensitivity estimates for the relevant signatures at Run~2 and~Run~3 of the LHC. We~also touch upon the sensitivity associated with hypothetical LHC searches for a $hh + E_{T, \rm miss}$ signal. 

\subsection{Searches for light charged Higgs bosons}
\label{sec:chargedHiggssearches}

\begin{figure}[t!]
\begin{center}
\includegraphics[width=0.475\textwidth]{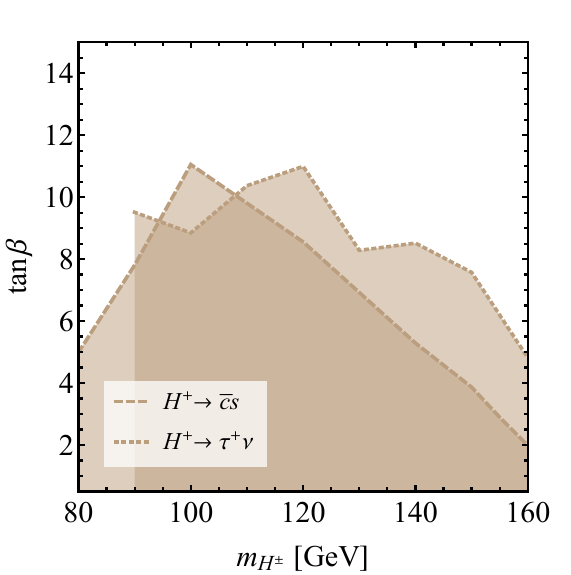}
\vspace{2mm}
\caption{\label{fig:chargedhiggs} Constraint on the $m_{H^\pm}\hspace{0.125mm}$--$\hspace{0.5mm}\tan \beta$ plane in the type-I~2HDM+$a$ model imposed by the search~\cite{CMS:2020osd} and~\cite{ATLAS:2018gfm} for light charged Higgs bosons in the $H^+ \to c \bar s$ and $H^+ \to \tau^+ \nu$ channel, respectively. The shaded regions are excluded at $95\%$~CL. Further explanations can be found in the main text.} 
\end{center}
\end{figure}

We begin our discussion with deriving the model-independent constraints on light charged Higgs bosons that stem from existing LHC searches. 2HDM+$a$ model realisations with charged Higgs mass of $m_{H^\pm} \leq m_t - m_b \simeq168 \, {\rm GeV}$ can be probed by searching for a light $H^\pm$ that decays dominantly via $H^+ \to c \bar s$ or $H^+ \to \tau^+ \nu$. These channels have been targeted by both ATLAS and CMS in $t \bar t$ production where light charged Higgs bosons can appear in the top decay $t \to H^+ b \to c \bar s b$~\cite{CMS:2015yvc,CMS:2020osd} or $t \to H^+ b \to \tau^+ \nu b$~\cite{ATLAS:2018gfm,CMS:2019bfg}. In particular, these analyses have established model-independent upper limits on the branching ratios ${\rm Br} \left (t \to H^+ b \right ) {\rm Br} \left (H^+ \to c \bar s \right )$ and ${\rm Br} \left (t \to H^+ b \right ) {\rm Br} \left (H^+ \to \tau^+ \nu \right )$, respectively. As~shown in~Figure~\ref{fig:chargedhiggs} in the type-I~2HDM+$a$ model the obtained upper limits lead to stringent lower bounds on $\tan \beta$ for charged Higgs masses with $80 \, {\rm GeV} \leq m_{H^\pm} \leq 160 \, {\rm GeV}$. These bounds have been obtained at the tree level. Notice~that the displayed constraints assume the alignment limit and that $m_{A,H,a} > 80 \ {\rm GeV}$ so that the charged Higgs cannot decay into a final state with a $W$ boson. Under these assumptions the depicted limits become independent of the remaining 2HDM+$a$ parameters that appear on the right-hand side in~(\ref{eq:2HDMainput}). Specifically, for $\tan \beta = 5$ the analysis~\cite{CMS:2020osd} imposes the $95\%$~CL lower limit of $m_{H^\pm} > 145 \, {\rm GeV}$ on the charged Higgs-boson mass, while the search~\cite{ATLAS:2018gfm} leads to $m_{H^\pm} > 160 \, {\rm GeV}$ at $95\%$~CL. In contrast for charged Higgs masses above the top-quark threshold the dominant decay mode of the charged Higgs boson is $H^+ \to t \bar b$. The latest LHC searches for the $H^+ \to t \bar b$ channel~\cite{CMS:2019rlz,ATLAS:2021upq} provide constraints that are too weak to test type-I~2HDM+$a$ scenarios with $\tan \beta =5$ even for $m_{H^\pm} = 200 \, {\rm GeV}$. We add that the charged Higgs-boson searches for $H^+ \to W^+ A \to W^+ \mu^+ \mu^-$~\cite{CMS:2019idx,ATLAS:2021xhq} and $H^+ \to W^+ H \to W^+ \tau^+ \tau^-$~\cite{CMS:2022jqc} do not pose any relevant restrictions on the parameter space of the type-I~2HDM+$a$ model studied below. This is related to the fact that the branching ratios ${\rm Br} \left ( A \to \mu^+ \mu^-\right )$ and ${\rm Br} \left ( H \to \tau^+ \tau^- \right)$ do not even reach a percent in the benchmarks considered in our work.

\subsection{Setup for truth-level sensitivity analyses}
\label{sec:signalbackground}

As far as the new-physics signals are concerned, the starting point of our MC simulations is a {\tt UFO} implementation~\cite{Degrande:2011ua} of the full 2HDM+$a$ model of type~I~as described in~Section~\ref{sec:2hdma}. Our implementation has been obtained by means of the {\tt FeynRules~2}~\cite{Alloul:2013bka} and~{\tt NLOCT}~\cite{Degrande:2014vpa} packages. The new {\tt UFO} can be downloaded from the public repository of the~LHC~DM~WG~\cite{UFO}. It~contains the general set of interactions of the type-I~2HDM+$a$ model, granting users the freedom to choose all input parameters listed on the right-hand side of~(\ref{eq:2HDMainput}) independently, with the exception of $\lambda_{P3}$. This quartic self-coupling has been omitted, since it is irrelevant for the collider phenomenology of the 2HDM+$a$ model.

All signal and most of the backgrounds are generated with {\tt MadGraph5\_aMC@NLO}~\cite{Alwall:2014hca}. The unstable final-state particles in the signal samples are decayed using~{\tt MadSpin}~\cite{Artoisenet:2012st}. Due to technical limitations in \texttt{MadSpin}~\cite{Hirschi:2015iia}, we neglect spin correlations. However, these would have a minimal impact on the results provided below. The $V + {\rm jets}$ backgrounds with $V = W, Z$ are generated using leading-order~(LO) matrix elements and contain up to three jets. The $V + {\rm jets}$ sample is merged using the CKKW-L procedure~\cite{Lonnblad:2001iq} and normalised to the inclusive cross section obtained at the next-to-next-to-leading order~(NNLO) in QCD~\cite{Anastasiou:2003ds}. The backgrounds corresponding to the $pp\to t\bar{t}$, $pp\to tW$, $pp\to t\bar{t}h$, and $pp\to Vh$ processes have been simulated using {\tt POWHEG~BOX}~\cite{Alioli:2010xd} with the inclusive cross sections normalised to their most precise available determinations. We use the NNLO plus next-to-next-to-leading logarithmic~(NNLL)~QCD prediction for~$t\bar{t}$~production~\cite{Czakon:2013goa,Czakon:2011xx}, the next-to-leading order~(NLO) plus NNLL QCD prediction for the $tW$ process~\cite{Kidonakis:2010ux}, NLO QCD plus NLO EW corrections for $t\bar{t}h$ production~\cite{LHCHiggsCrossSectionWorkingGroup:2016ypw}, the NNLO QCD computation of the $q\bar{q}\to Vh$ channels~\cite{Campbell:2016jau}, and NNLO plus NNLL QCD as well as NLO EW corrections for the $gg\to Vh$ processes~\cite{Brein:2012ne,Harlander:2014wda}. The rest of the background samples are generated with {\tt MadGraph5\_aMC@NLO} using NLO matrix elements, with their inclusive cross section taken from the values provided by the MC~generator. All signal and background samples use the {\tt NNPDF3.0}~\cite{NNPDF:2014otw} parton distribution functions for the calculation of the partonic cross sections. For showering and hadronisation of all samples, {\tt PYTHIA~8.2}~\cite{Sjostrand:2014zea} is used with the A14~tune~\cite{TheATLAScollaboration:2014rfk}. 

The truth-level analyses described in the following sections are implemented in the {\tt RIVET} framework~\cite{Buckley:2010ar,Bierlich:2019rhm} using the smearing functionality developed in~\cite{Buckley:2019stt} to simulate detector-level effects. In our analyses, charged leptons,~i.e.~$\ell = e, \mu$, are required to have a transverse momentum of $p_{T, \ell}>7 \, {\rm GeV}$, a pseudorapidity of $|\eta_\ell|<2.5$, and are dressed with photons that lie within a separation of $\Delta R<0.1$ from the lepton. Jets are reconstructed from stable particles located within $|\eta|<5$, excluding the previously mentioned charged leptons but including non-prompt muons and neutrinos originating from hadron decays. The jet reconstruction is performed with the {\tt FastJet}~\cite{Cacciari:2011ma} package, using the anti-$k_t$ algorithm~\cite{Cacciari:2008gp} with a radius parameter of $R=0.4$. Jets that have at least one ghost-associated $B$-hadron~\cite{Cacciari:2008gn} with $p_{T,B} >5 \, {\rm GeV}$ are identified as $b$-jets.

The sensitivity calculation for each search relies on the asymptotic approximation of the profile-likelihood test statistic~\cite{Cowan:2010js} corresponding to a $95\%$~CL. This~calculation is conducted in bins of a discriminating variable, with the bin width chosen appropriately to maximise the significance while minimising the impact of statistical fluctuations arising from the limited statistics of the simulated event samples.

\subsection[Search for $h + E_{T, \rm miss}$ final state]{Search for $\bm{h + E_{T, \rm miss}}$ final state}
\label{sec:hMETsearch}

To derive constraints on type-I~2HDM+a model realisations that lead to an important $h + E_{T, \rm miss}$ signature, we develop our own truth-level analysis that targets the $h \to b \bar b$ decay, based on the search strategy detailed in~\cite{ATLAS:2023zkt}. We stress that we include only the $gg \to h \chi \bar \chi$ but not the $gg \to h 4 \chi$ contribution in our signal generation of the $h + E_{T, \rm miss}$ final state. Relevant graphs for the former channel can be found in~Figure~\ref{fig:diagrams1}. We select events with no charged leptons as defined in~Section~\ref{sec:signalbackground}. Events are required to contain no $\tau$ leptons and to satisfy $E_{T, \rm miss} > 150 \,{\rm GeV}$. To account for the inefficiency of the $E_{T, \rm miss}$ trigger at such $E_{T, \rm miss}$ values, a trigger requirement is incorporated in the {\tt RIVET} routine using the efficiency parametrisation provided in~\cite {ATLAS:2021shl}. Jets are required to satisfy $p_{T, j} > 20 \, {\rm GeV}$ and $|\eta_j|<2.5$. Only events with exactly two $b$-jets and less than six jets are considered. The~$b$-jet pair is required to satisfy $m_{b \bar b} > 50 \, {\rm GeV}$ and $\Delta R_{b\bar{b}} < 3.3$. Events are further required to have $E_{T, \rm miss}/\sqrt{H_T}>10$ GeV$^{1/2}$, where $H_T$ is the scalar sum of the $p_T$ of all jets in the event. This cut mimics the ``$E_{T, \rm miss}$ significance'' cut that is used in~\cite{ATLAS:2023zkt}. To~reduce contamination from events with fake $E_{T, \rm miss}$, events containing any jet with $\Delta \phi (\vec{p}_{T, \rm miss}, \vec{p}_{T,j})<\pi/10$ are~rejected. Here $\vec{p}_{T, \rm miss}$~$(\vec{p}_{T,j}$) denotes the relevant three-vector with magnitude $E_{T, \rm miss}$~($p_{T, j}$). Finally, the $t \bar t$ background contribution in events with no leptons is suppressed by cutting on two top-quark mass proxy variables
\beq \label{eq:mtnearfar}
m_{\rm top}^{{\rm near}/{\rm far}} = \sqrt{2 \hspace{0.25mm} p_{T, b_{{\rm near}/{\rm far}}} \hspace{0.25mm} E_{T, \rm miss} \left [ 1 - \cos \Delta \phi (\vec{p}_{T, b_{{\rm near}/{\rm far}}}, \vec{p}_{T, \rm miss} ) \right ]} \,, 
\eeq
where near (far) refers to the $b$-jet that is nearer to (farther from) the $\vec{p}_{T, \rm miss}$ three-vector in azimuthal angle $\phi$. The $b$-jet closer to (farther from) $\vec{p}_{T, \rm miss}$ is used to calculate $m_{\rm top}^{\rm near}$~($m_{\rm top}^{\rm far}$). Events are retained only if they satisfy $m_{\rm top}^{\rm near} > 180 \, {\rm GeV}$ and $m_{\rm top}^{\rm far} > 200\, {\rm GeV}$.

We note that while the actual ATLAS search~\cite{ATLAS:2023zkt} defines a signal region (SR) by implementing a cut on the reconstructed mass of the Higgs candidate, $m_{b \bar b}$, and subsequently utilises the transverse mass of the $A$ candidate, $m_{T} (b\bar{b}, E_{T, \rm miss})$, as a discriminant, we simplify this procedure. Instead, following~\cite{ATLAS:2021shl}, we estimate the sensitivity using the~$m_{b\bar{b}}$ distribution to estimate the sensitivity. As demonstrated in~Section~\ref{sec:constraintssummary}, our simplified approach nonetheless leads to exclusions even with the LHC~Run~2 dataset. A~more sophisticated analysis using $m_{T} (b\bar{b}, E_{T, \rm miss})$ in bins of $m_{b\bar{b}}$, akin to that performed in the work~\cite{ATLAS:2023zkt}, is, however, expected to provide even better exclusion limits.

\subsection[Search for $b \bar b + Z$ signatures]{Search for $\bm{b \bar b + Z}$ signatures}
\label{sec:ZHsearches}

In the case of the $b \bar b + Z$ signal arising for instance from the right diagram in~Figure~\ref{fig:diagrams4}, our truth-level analysis follows the search for $pp \to A\to ZH$ production described in~\cite{ATLAS:2020gxx}. We target the $\ell^+ \ell^- b \bar b$ channel by selecting only events that contain exactly two electrons or muons of opposite charge with $80 \, {\rm GeV} < m_{\ell^+ \ell^-} < 100 \, {\rm GeV}$. The leading and subleading lepton are required to have $p_{T, \ell} >27 \, {\rm GeV}$ and $p_{T, \ell} > 7 \, {\rm GeV}$, respectively. Reconstructed~jets have to satisfy $p_{T, j} > 25 \, {\rm GeV}$ and $|\eta_j| < 2.5$. Only events with exactly two $b$-jets are retained. For hypotheses with $m_H > 150 \, {\rm GeV}$, the $p_T$ of the leading $b$-jet has to satisfy $p_{T, b} > 45 \, {\rm GeV}$ in addition. The events are further required to have $E_{T,\rm miss}/\sqrt{H_T}<3.5$ GeV$^{1/2}$ and to satisfy $\sqrt{\sum_{i=\ell,b}{p_{T,i}^2}}/m_{\ell^+\ell^- b\bar{b}}>0.3 \, (0.4)$, when testing hypotheses with $m_H<200 \, {\rm GeV}$ ($m_H \geq 200 \, {\rm GeV}$). Here $m_{\ell^+\ell^- b\bar{b}}$ denotes the invariant mass of the $\ell^+\ell^- b\bar{b}$ system. 

After imposing the above requirements, an SR is defined by selecting events for which the reconstructed invariant mass $m_{b \bar b}$ of the two $b$-jets falls within a certain range from the~$m_H$~hypothesis. Specifically, the SR is defined by the condition $m_H - 20 \, {\rm GeV} < m_{b \bar b} < m_H + 20 \, {\rm GeV}$ for hypotheses with $m_H <100 \, {\rm GeV}$, and $0.85 \hspace{0.5mm} m_H - 20 \, {\rm GeV} < m_{b \bar b} < m_H + 20 \, {\rm GeV}$ for hypotheses with $m_H \geq 100 \, {\rm GeV}$. Notice that the mass window is enlarged when testing higher~$m_H$ values to take into account the worsening of the experimental $m_{b \bar b}$ resolution. For events that fulfil the SR requirements, the invariant mass $m_{\ell^+\ell^- b\bar{b}}$ of the $A$ candidate is reconstructed from the four-momenta of the two leptons and the two $b$-jets. The significance is then calculated in bins of the $m_{\ell^+\ell^- b\bar{b}}$ distribution. In order to validate our truth-level analysis, we have performed a recast of the ATLAS search~\cite{ATLAS:2020gxx}. This was found to give a similar exclusion compared to our truth-level analysis in the region with $m_H \geq 130 \, {\rm GeV}$. However, since \cite{ATLAS:2020gxx} only considers signals with $m_H \geq 130 \, {\rm GeV}$, the region with lower $m_H$ cannot be accessed via a recast. The expected exclusion obtained by our truth-level analysis, which instead covers the entire accessible $m_H$ region down to $m_H = 50 \,{\rm GeV}$, is shown in~Figure~\ref{fig:exA}.

We add that the $b \bar b + Z$ process can also be searched for in the $\nu \bar \nu b \bar b$ channel. This~final state has been studied for example by ATLAS in~\cite{ATLAS:2023zkt}. Recasting these limits, we find that for the type-I 2HDM+$a$ benchmark models proposed in Section~\ref{sec:benchmarks}, the constraints on the parameter space resulting from the $b \bar b \nu \bar \nu$ search~\cite{ATLAS:2023zkt} are always slightly weaker than those from our own $\ell^+ \ell^- b \bar b$ truth-level analysis. Therefore, the restrictions from a recast of \cite{ATLAS:2023zkt} are not included in~Section~\ref{sec:constraintssummary} when summarising the LHC sensitivity. Similar considerations apply to the $t \bar t + Z$ channel, investigated in the $t \bar t \ell^+ \ell^-$ final state in~\cite{ATLAS:2023zkt}. Upon reinterpreting these results, we observe that the latter search does not impose any restrictions on the type-I 2HDM+$a$ benchmark models considered in this paper due to the suppression of the production cross section for $\tan\beta = 5$.

\subsection[Search for $b \bar b + E_{T, \rm miss}$ final states]{Search for $\bm{b \bar b + E_{T, \rm miss}}$ final states}
\label{sec:HMETsearches}

Our $b \bar b + E_{T, \rm miss}$ truth-level analysis targets events with a pair of $b$-jets of invariant mass $m_{b\bar{b}}\neq 125 \, {\rm GeV}$. Notice that in the context of type-I~2HDM+$a$ model realisations with a non-degenerate sector of $A$, $H$, and $H^\pm$ states, such a signal can arise resonantly from both $gg \to A \to Ha \to b \bar b \chi \bar \chi$ and $gg \to A \to ZH \to \nu \bar \nu b \bar b$ production. The corresponding Feynman diagrams are shown on the left in~Figure~\ref{fig:diagrams5} and on the right in~Figure~\ref{fig:diagrams4}, respectively. While a recast of the ATLAS~searches~\cite{ATLAS:2021shl,ATLAS:2023zkt} can provide constraints for $m_H \geq 125 \, {\rm GeV}$, we employ in this article the search strategy detailed in~Section~\ref{sec:hMETsearch} to derive the LHC sensitivity to the $b \bar b + E_{T, \rm miss}$ signature. As shown in~Figure~\ref{fig:exA}, in this~way, we are able to extend the mass reach of the $b \bar b + E_{T, \rm miss}$ search from $m_H \geq 125 \, {\rm GeV}$ down to $m_H = 50 \, {\rm GeV}$. We stress that both signal processes,~i.e.~$b \bar b \chi \bar \chi$ and $\nu \bar \nu b \bar b$, are combined to derive the sensitivity of our $b \bar b + E_{T, \rm miss}$ truth-level analysis.

We add that the light $H$ states targeted by our search strategy decay predominantly to bottom-quark pairs. In the considered benchmarks, typical values for the branching ratios of interest are ${\rm Br} \left (H \to b \bar b \right ) \simeq 80\%$, ${\rm Br} \left (H \to \tau^+  \tau^- \right ) \simeq 10\%$, and ${\rm Br} \left (H \to \gamma \gamma \right ) \simeq 10^{-4}$. Comparing these numbers with those of the $125 \, {\rm GeV}$ Higgs boson,~i.e.~${\rm Br} \left (h \to b \bar b \right ) \simeq 58\%$, ${\rm Br} \left (h \to W^+ W^- \right ) \simeq 21\%$, ${\rm Br} \left (h \to \tau^+ \tau^- \right ) \simeq 6 \%$ and ${\rm Br} \left (h \to \gamma \gamma \right ) \simeq 2 \cdot 10^{-3}$, one concludes that search strategies that would target $H \to \gamma \gamma$ instead of $H \to b \bar b$ are expected to provide a weaker sensitivity than the truth-level analysis described above. A search strategy focusing on the $H \to \tau^+ \tau^-$ decay could present a complementary alternative to our $b \bar b + E_{T, \text{miss}}$ analysis. Such an analysis has been performed in the context of the type-II 2HDM+$a$ model by ATLAS in~\cite{ATLAS:2023ild}.

\subsection[Search for $Z Z + E_{T, \rm miss}$ signals]{Search for $\bm{ZZ + E_{T, \rm miss}}$ signals}
\label{sec:ZZMETsearches}

Searches for a $Z Z + E_{T, \rm miss}$ signature have been performed by ATLAS in~\cite{ATLAS:2021wob,ATLAS:2024bzr}. The~former search considers events with three or four leptons, employing seven distinct SRs and using the invariant mass of the four leptons, $m_{4 \ell}$, as a discriminant. The former analysis is more model-independent, using the total yield measured in 22 distinct SRs as a discriminant and is therefore easier to recast. Notice that a $Z Z + E_{T, \rm miss}$ signal can arise in the 2HDM+$a$ model of type I with non-degenerate BSM Higgs bosons from the right graph in~Figure~\ref{fig:diagrams5}. This contribution gives rise to events with four leptons, a large amount of~$E_{T, \rm miss}$ and relatively small $m_{4\ell}$ values. 

Since recasting~\cite{ATLAS:2024bzr} using the 2HDM+$a$ signal has been found not to provide an exclusion for any of the considered type-I benchmarks, we have developed our own truth-level analysis. While our dedicated search strategy is based on the selection criteria employed in~\cite{ATLAS:2024bzr}, it uses the transverse mass of the $A$ candidate, $m_T (4\ell, E_{T,\rm miss})$, instead of $m_{4\ell}$ as a discriminant. Specifically, final states with exactly four leptons all with $p_{T, \ell}>25 \, {\rm GeV}$ and $|\eta_\ell|<2.5$ are selected. Events are required to have at least one lepton pair compatible with a $Z$-boson candidate,~i.e.~two leptons of the same flavour, of opposite charge, and with an invariant mass satisfying $|m_{\ell^+\ell^-}-m_Z|<10 \, {\rm GeV}$. Following~\cite{ATLAS:2024bzr}, events are separated into bins of $E_{T,\rm miss}$ and~$m_{4\ell}$. The generated signals have been found to populate almost exclusively the region with $E_{T, \rm miss}>50 \, {\rm GeV}$ and $m_{4\ell} < 400 \, {\rm GeV}$, and therefore to simplify the analysis, only events in this SR are retained for the sensitivity calculation. The sensitivity is evaluated in bins of the $m_T (4\ell, E_{T,\rm miss})$ distribution, which offers a higher signal-to-background separation compared to the $m_{4\ell}$ distribution used in~\cite {ATLAS:2021wob,ATLAS:2024bzr}.

\subsection{Summary of constraints}
\label{sec:constraintssummary}

Based on the analysis strategies outlined in~Sections~\ref{sec:hMETsearch}, \ref{sec:ZHsearches}, \ref{sec:HMETsearches}, and \ref{sec:ZZMETsearches}, we will now present sensitivity estimates for the $h + E_{T, \rm miss}$, $b\bar b + Z$, $b \bar b + E_{T, \rm miss}$, and $ZZ + E_{T, \rm miss}$ signals. Additionally, we will consider the constraints that arise from searches for light charged Higgs bosons discussed in~Section~\ref{sec:chargedHiggssearches} and explore the impact of EW precision observables and theoretical restrictions on the type-I~2HDM+$a$ parameter space. 

\begin{figure}[t!]
\begin{center}
\includegraphics[width=0.99\textwidth]{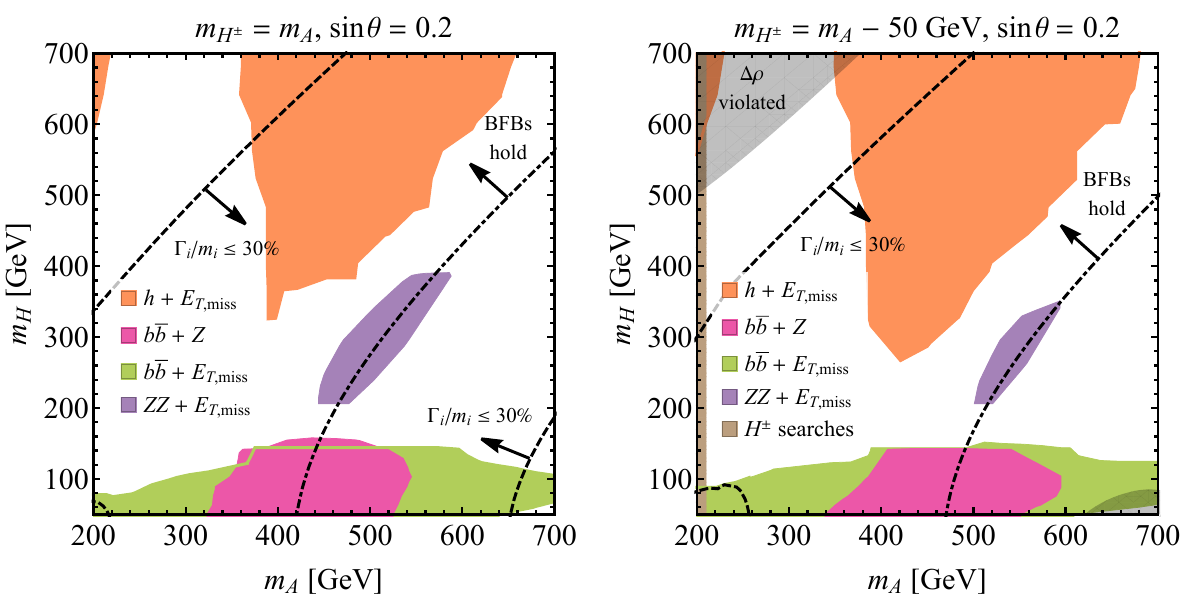}

\includegraphics[width=0.99\textwidth]{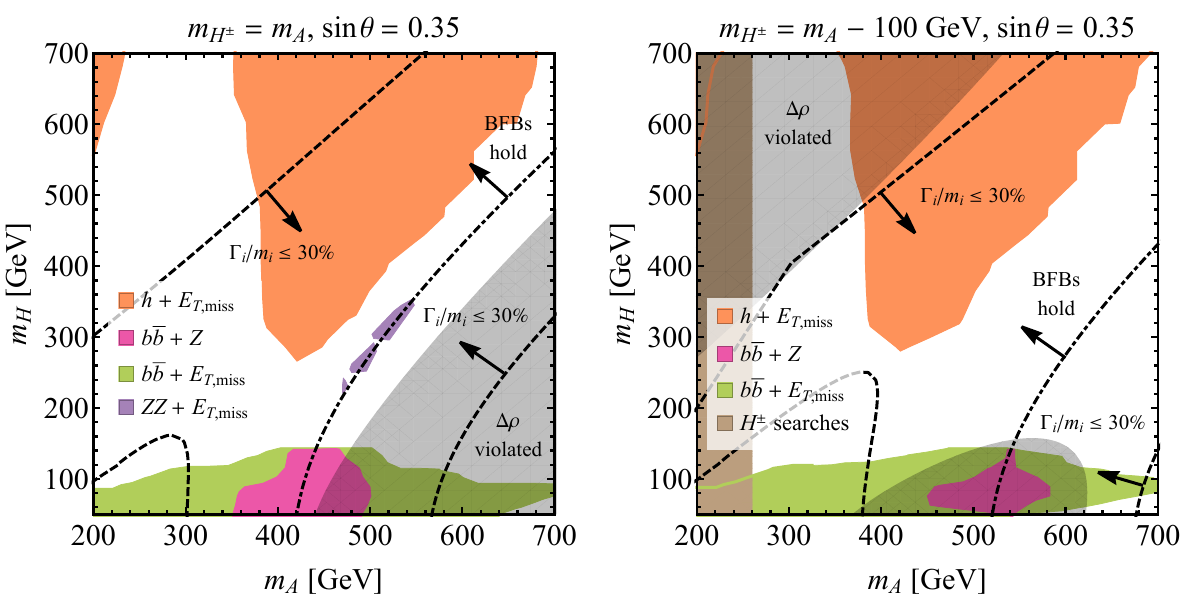}
\vspace{0mm}
\caption{\label{fig:exA} Constraints on the type-I~2HDM+$a$ model resulting from hypothetical searches for $h + E_{T, \rm miss}$, $b\bar b + Z$, $b \bar b + E_{T, \rm miss}$, and $ZZ + E_{T, \rm miss}$ final states. The different plots correspond to four distinct choices of $m_{H^\pm}$ and $\sin \theta$ as indicated in the panel headings. The~remaining parameters are set to~(\ref{eq:generalparameter}). The limits on the charged Higgs-boson mass derived from~\cite{CMS:2020osd,ATLAS:2018gfm} are also shown. Additionally, regions of the $m_A\hspace{0.125mm}$--$\hspace{0.5mm}m_H$~planes that are in conflict with EW precision measurements, specifically the bound~(\ref{eq:rhoconstraint}), are depicted in grey. Finally, the parameter regions leading to vacuum stability, as encoded by the BFB conditions~(\ref{eq:BFB}), and the requirements $\Gamma_i/m_i \leq 30\%$ on the relative total decay widths of all the BSM spin-$0$ states $i = A, H, H^\pm, a$ are overlaid as dashed-dotted and dashed lines, respectively. See main text for additional details.} 
\end{center}
\end{figure}

The outcomes of our LHC sensitivity studies are presented in~Figure~\ref{fig:exA}. The depicted~$m_A\hspace{0.125mm}$--$\hspace{0.5mm}m_H$ scans represent four different choices of~$m_{H^\pm}$ and~$\sin \theta$. Parameters not explicitly indicated in the panel headings are set according to~(\ref{eq:generalparameter}). Consequently, the analysed benchmarks align with those explored in~Figure~\ref{fig:muA}. In the case of the $h + E_{T, \rm miss}$, the $b\bar b + Z$, and the $b \bar b + E_{T, \rm miss}$ searches, the shown 95\%~CL exclusions correspond to an integrated luminosity of $140 \, {\rm fb}^{-1}$. Instead, for the $ZZ + E_{T, \rm miss}$ channel, given its weaker sensitivity, we have assumed $400 \, {\rm fb}^{-1}$ of $13 \, {\rm TeV}$ LHC data to set the constraints. In addition to the $h + E_{T, \rm miss}$, $b\bar b + Z$, $b \bar b + E_{T, \rm miss}$, and $ZZ + E_{T, \rm miss}$ constraints, we have also included in the plots the limit $m_{H^\pm} > 160 \, {\rm GeV}$ on the charged Higgs-boson mass that derives from the searches~\cite{CMS:2020osd,ATLAS:2018gfm} in the case of $\tan \beta = 5$ (cf.~Section~\ref{sec:chargedHiggssearches} for details). The~parts of the 2D planes that are disfavoured by the bound (\ref{eq:rhoconstraint}) on the $\rho$ parameter are indicated as~well. Finally, the theoretically favoured parameter regions following from vacuum stability, as encoded by the BFB conditions~(\ref{eq:BFB}), and the requirements $\Gamma_i/m_i \leq 30\%$ on the relative total decay widths of all the BSM spin-$0$ states $i = A, H, H^\pm, a$ are displayed. Given~the indicative nature of the latter two constraints the corresponding parameter regions are depicted by lines rather than shaded areas. 

The results show in the four panels of~Figure~\ref{fig:exA} reveal that the considered searches, namely $h + E_{T, \rm miss}$, $b\bar b + Z$, $b \bar b + E_{T, \rm miss}$, and $ZZ + E_{T, \rm miss}$, enable the exploration of complementary regions in the $m_A\hspace{0.125mm}$--$\hspace{0.5mm}m_H$ planes. As anticipated from~Figure~\ref{fig:muA}, the probed regions remain largely independent of the specific choices of $m_{H^\pm}$ and $\sin \theta$. Regarding the displayed $h + E_{T, \rm miss}$ constraints, two remarks seem important. First, one observes that the bulk region of all $h + E_{T, \rm miss}$ constraints has a sharp lower cut-off within the mass range $350 \, {\rm GeV} \lesssim m_A \lesssim 400 \, {\rm GeV}$. This~feature is readily understood by recalling that in the 2HDM+$a$ model the $E_{T, \rm miss}$ spectrum in the mono-Higgs channel has a Jacobian peak with an endpoint determined by the kinematics of the $A \to ha$ process~\cite{No:2015xqa,Bauer:2017ota}. The presence of the Jacobian peak allows to derive the following approximate inequality: 
\beq \label{eq:mAinequality}
m_A \gtrsim \sqrt{m_h^2+ m_a^2 + 2 \left[ \big ( E_{T, \rm miss}^{\rm cut} \big)^2 + \sqrt{\left ( m_h^2 + \big ( E_{T, \rm miss}^{\rm cut} \big)^2 \right ) \left ( m_a^2 + \big ( E_{T, \rm miss}^{\rm cut} \big)^2 \right )} \right ] } \,.
\eeq 
This inequality provides an estimate for the lower limit on $m_A$ for which a given mono-Higgs search is expected to be sensitive. Here $E_{T, \rm miss}^{\rm min}$ denotes a minimal missing transverse energy requirement,~i.e.~$E_{T, \rm miss} > E_{T, \rm miss}^{\rm min}$. Such a condition is employed in all existing mono-Higgs analyses \cite{CMS:2018zjv,ATLAS:2019wdu,ATLAS:2021jbf,ATLAS:2021shl,CMS:2022sfl,ATLAS:2023ild} and is imposed by the high $E_{T, \text{miss}}$ requirement applied in the~$E_{T, \rm miss}$ trigger that these analyses rely on. Notably,~utilising $m_h = 125 \, {\rm GeV}$, $m_a = 100 \, {\rm GeV}$, and $E_{T, \rm miss}^{\rm min} = 150 \, {\rm GeV}$ in (\ref{eq:mAinequality}) implies $m_A \gtrsim 375 \, {\rm GeV}$, which adequately accounts for the sharp lower cut-offs observed in the~$h + E_{T, \rm miss}$ constraints depicted in~Figure~\ref{fig:exA}. Notice that $E_{T, \rm miss}^{\rm min} = 150 \, {\rm GeV}$ corresponds to the minimal $E_{T, \rm miss}$ selection imposed in our $h + E_{T, \rm miss}$ search as described in~Section~\ref{sec:hMETsearch}. The above discussion makes clear that the sensitivity of $h + E_{T, \rm miss}$ searches to the parameter space of 2HDM+$a$ models with $m_a \simeq 100 \, {\rm GeV}$ and $m_A \lesssim 300 \, {\rm GeV}$ would require to push the minimal $E_{T, \rm miss}$ requirement down to $E_{T, \rm miss}^{\rm min} \simeq 100 \, {\rm GeV}$. It would be interesting to investigate whether this could be achieved by triggering on $b$-jets \cite{ATLAS:2021piz}, as done for example in~\cite {ATLAS:2022ygn}. Also note that in parameter regions where $m_H > m_A - m_a$ and for not too large values of $m_A$, the dominant decay mode of the CP-odd 2HDM pseudoscalar is $A \to ha$. This feature, combined with the fact that ${\rm Br} \left (A \to ha \right)$ decreases while ${\rm Br} \left (A \to t \bar t \right)$ increases with increasing~$m_A$, determines the shape of the right edges of the bulk regions of the $h + E_{T, \rm miss}$ constraints. A~second observation is that in all panels, there exists a small region of parameter space around $m_A \simeq 200 \, {\rm GeV}$ and $m_H \simeq 650 \, {\rm GeV}$ that is excluded by our $h + E_{T, \rm miss}$ sensitivity study. Within this region, the $h + E_{T, \rm miss}$ signature emerges in the studied type-I~2HDM+$a$ benchmarks due to the box contribution to $gg \to h a \to b \bar b \chi \bar \chi$ --- cf.~the right diagram in~Figure~\ref{fig:diagrams1} --- and off-shell $gg \to a^\ast \to h a \to b \bar b \chi \bar \chi$ production. We add that our signal samples only contain the $gg \to h \chi \bar{\chi}$ contribution to the $h+E_{T, \rm miss}$ signal. The possible impact of the $gg \to h \chi\bar{\chi}\chi\bar{\chi}$ contribution is studied in~Appendix~\ref{app:h4chi}, where it is shown that the inclusion of the latter would enhance the sensitivity of our truth-level analysis slightly. We~finally mention that model realisations in the left upper corners in the panels of~Figure~\ref{fig:exA} lead to $\Gamma_H/m_H > 30\%$. However, since the $gg \to h \chi \bar{\chi}$ channel does not receive contributions from diagrams involving the exchange of an $H$, our simulation of the mono-Higgs signature does not suffer from the model dependence related to the precise treatment of large-width resonances.

Regarding the $b\bar b + Z$ and $b \bar b + E_{T, \rm miss}$ constraints in~Figure \ref{fig:exA}, it is evident that these searches can only probe model realisations with light CP-even Higgs bosons~$H$~in the mass range $50 \, {\rm GeV} \leq m_H \lesssim 150 \, {\rm GeV}$. Indeed, this observation nicely aligns with the findings in~Figure~\ref{fig:muA}, which indicate that for $m_H \gtrsim 200 \, {\rm GeV}$ the $b\bar b + Z$ and $b \bar b + E_{T, \rm miss}$ signatures are typically not the dominant final-state configurations. We note that for $m_H = 50 \, {\rm GeV}$, $m_a = 100 \, {\rm GeV}$, and $E_{T, \rm miss}^{\rm min} = 150 \, {\rm GeV}$, an inequality analogous to (\ref{eq:mAinequality}) implies that our $b \bar b + E_{T, \rm miss}$ search strategy is sensitive to resonant $gg \to A \to Ha \to b \bar b \chi \bar \chi$ production only if $m_A \gtrsim 340 \, {\rm GeV}$. Below $m_A \simeq 340 \, {\rm GeV}$, the $b \bar b + E_{T, \rm miss}$ exclusion is thus determined by the size of the box contribution to $gg \to H a \to b \bar b \chi \bar \chi$, $gg \to a^\ast \to H a \to b \bar b \chi \bar \chi$ production and the $gg \to A \to Z H \to \nu \bar \nu b \bar b$ channel. We also note that due to the $E_{T, \rm miss}^{\rm min} = 150 \, {\rm GeV}$ requirement imposed in our $b \bar b + E_{T, \rm miss}$ search, the wedge in the $m_A\hspace{0.125mm}$--$\hspace{0.5mm}m_H$ planes with $100 \, {\rm GeV} \lesssim m_A-m_H\lesssim 200 \, {\rm GeV}$ cannot be probed.

In~comparison to the other searches, our analysis targeting the $ZZ + E_{T, \rm miss}$ final state exhibits weaker sensitivity, yielding constraints only upon application to the entire dataset expected to be available at the end of LHC~Run~3. With an integrated luminosity of $400 \, {\rm fb}^{-1}$, searches focusing on the $ZZ + E_{T, \rm miss}$ channel should, however, exhibit sensitivity to the parameter space centred around $m_A \simeq 500 \, {\rm GeV}$ and $m_H \simeq 300 \, {\rm GeV}$ in three out of the four studied benchmarks. This aligns with the findings in~Figure~\ref{fig:muA}.

\subsection[Comments on other $E_{T, \rm miss}$ channels]{Comments on other $\bm{E_{T, \rm miss}}$ channels}
\label{sec:others}

From the panels in~Figure \ref{fig:exA}, it is apparent that the analysed constraints from $h + E_{T, \rm miss}$, $b\bar b + Z$, $b \bar b + E_{T, \rm miss}$, $ZZ + E_{T, \rm miss}$, and charged Higgs searches are not sufficient to probe the entire $m_A\hspace{0.125mm}$--$\hspace{0.5mm}m_H$ planes. This prompts the obvious question: what search strategies could possibly explore the uncharted parameter space in the benchmarks studied above? One~candidate to test some type-I~2HDM+$a$ model realisations with $m_A \lesssim 375 \, {\rm GeV}$ and $m_H \gtrsim 150 \, {\rm GeV}$ is the $Z+E_{T, \rm miss}$ channel. This follows from the simple observation that the minimal $E_{T, \rm miss}$ requirement in $Z+E_{T, \rm miss}$ searches like~\cite{ATLAS:2021gcn} is only $E_{T, \rm miss}^{\rm cut} = 90 \, {\rm GeV}$. Using~(\ref{eq:mAinequality}) appropriately modified to capture the kinematics of the $A \to ZH$ process one obtains $m_A \gtrsim 300 \, {\rm GeV}$ for $m_Z \simeq 91.2 \, {\rm GeV}$, $m_H \simeq 150 \, {\rm GeV}$, and $E_{T, \rm miss}^{\rm cut} = 90 \, {\rm GeV}$. Together with the fact that the $Z + E_{T, \text{miss}}$ signal in the benchmark scenarios studied in~Figure~\ref{fig:exA} receives contributions from various channels such as $gg \to H \to Z a$ (cf.~the right graph in~Figure~\ref{fig:diagrams6}), $gg \to A \to Z H \to Z aa$, and $gg \to A \to H a \to Z aa$ (as illustrated in~Figure~\ref{fig:diagrams2}), suggests that future searches for the $Z + E_{T, \text{miss}}$ final state could indeed offer some sensitivity to the parameter space where $m_A \lesssim 375 \, {\rm GeV}$. Due to the intricate nature of the contributions to the $Z + E_{T, \rm miss}$ signature within the examined type-I 2HDM+$a$ benchmarks, we defer a sensitivity study to future research. Looking at~Figure~\ref{fig:muA} and bearing in mind that for sufficiently large values of $m_A$ the dominant decay of the 2HDM pseudoscalar is $A \to t \bar t$, searches for final states such as~$t \bar t$~\cite{ATLAS:2017snw,CMS:2019pzc,ATLAS-CONF-2024-001} or $4t$~\cite{CMS:2019rvj,ATLAS:2020hpj,ATLAS:2021kqb,CMS:2023zdh,ATLAS:2023ajo} could in the future provide sensitivity to the parameter space to the right of the bulk regions excluded by our $h + E_{T, \rm miss}$ analysis~strategy. 

Let us now turn our attention to other novel signatures that emerge in the context of the type-I 2HDM+$a$ model, which we have not explored in detail. In this case one observes that in the parameter space centred around $m_A \simeq 250 \, {\rm GeV}$ and $m_H \simeq 500 \, {\rm GeV}$, the~$hh + E_{T, \rm miss}$ signal can emerge as the predominant final state. While a detailed sensitivity study of $hh + E_{T, \rm miss}$ is beyond the scope of this work, one can use the information provided in~\cite{Blanke:2019hpe}, which studies the $4b + E_{T, \rm miss}$ final state, to get a naive estimate of the potential LHC reach. Considering all four benchmark models in~Figure~\ref{fig:muH} and assuming $m_A \simeq 225 \, {\rm GeV}$ and $m_H \simeq 500 \, {\rm GeV}$, the cross section of $gg \to H \to AA \to haha \to 4b 4\chi$ production falls within the range of $[3, 26] \, {\rm fb}$. The discovery cross section at the high-luminosity option of the LHC~(HL-LHC) with $3000 \, {\rm fb}^{-1}$ is estimated to be around~$30 \, {\rm fb}$ for this mass configuration~\cite{Blanke:2019hpe}. This suggests that future $hh + E_{T, \rm miss}$ searches at the LHC should exhibit some sensitivity to the benchmark models studied in~Figure \ref{fig:muH}. In~fact, the final states $b \bar b \gamma \gamma + E_{T, \rm miss}$ and $b \bar b W W^\ast + E_{T, \rm miss}$ have been shown~\cite{Flores:2019hcf} to be also quite promising for detecting a $hh + E_{T, \rm miss}$ signal at the HL-LHC. These observations lead us to believe that comprehensive studies of the $hh + E_{T, \rm miss}$ channel are highly worthwhile in the context of the type-I~2HDM+$a$ model.

\section{Comments on relic density and direct detection predictions}
\label{sec:relicdensity}

The relic density obtained in the type-I 2HDM+$a$ model exhibits very similar characteristics to the~relic density predicted in the type-II 2HDM+$a$ model with degenerate BSM Higgs bosons, which has been extensively studied in~\cite{LHCDarkMatterWorkingGroup:2018ufk,Arcadi:2024ukq} --- see also~\cite{Argyropoulos:2022ezr}. A common feature across all benchmarks considered in our work is that the relic density is consistently overabundant for $m_a > 2 m_{\chi}$,~i.e.~when the decay $a \to \chi \bar \chi$ is kinematically allowed. Given~that the $E_{T,\rm miss}$ signatures examined earlier are produced by the $a \to \chi\bar{\chi}$ decay, it follows that identifying benchmarks where $E_{T,\rm miss}$ signals are predominant while also achieving the correct relic density is difficult.
 
A depletion of the relic density is achieved for $m_a \simeq 2m_\chi$ through the resonant $\chi\bar{\chi}\to a\to f\bar{f}$ process. Furthermore, DM annihilation via $ha$, $hA$, $t\bar{t}$, $Zh$, and $ZH$ into SM final states can also contribute to decreasing the relic density. Consequently, in the four benchmarks presented in~Figure~\ref{fig:exA}, regions can be found where the relic density is underabundant. We also note that, because of this characteristic, searches that do not involve the production of DM particles, such as the $b\bar{b}+Z$ or~$t\bar{t}$ channels, can always probe a region where the observed~relic density is reproduced, since one can always tune $m_{\chi}$ without affecting the signal cross section of the collider process.

In the 2HDM+$a$ model the DM-nucleon cross section that arises from tree-level exchange of the pseudoscalar $a$ is both momentum-suppressed and spin-dependent. However, a spin-independent (SI) DM-nucleon cross section arises once loop corrections are considered~\cite{Arcadi:2024ukq,Arcadi:2017wqi,Bell:2018zra,Abe:2018emu,Ertas:2019dew}. The SI DM-nucleon cross section can be approximated by
\beq \label{eq:sigmaSI}
\sigma_{\rm SI} \simeq \left ( \frac{m_N \hspace{0.5mm} m_\chi}{m_N + m_\chi} \right )^2  \frac{c_N^2}{\pi} \,,
\eeq
with $m_N \simeq 939 \, {\rm MeV}$ the average of the nucleon mass and $c_N$ the Wilson coefficient of the dimension-six DM-nucleon operator $O_N = \bar \chi \chi \bar N N$. As explained in~\cite{Abe:2018emu,Ertas:2019dew}, $c_N$~in~general receives contributions from Higgs-induced one-loop triangle and box diagrams, as well as two-loop contributions resulting in effective DM-gluon interactions. By~employing the formulae presented in~\cite{Abe:2018emu}, we find that for type-I~2HDM+$a$ realisations like~(\ref{eq:generalparameter}), the effects of one-loop box diagrams and two-loop graphs are numerically small. This is readily understood by realising that these contributions involve the couplings $g_{aq \bar q}$ and/or~$g_{Aq \bar q}$, which all scale as $\cot \beta$~$\big($cf.~(\ref{eq:2HDMfermions})$\big)$, leading to a suppression of these corrections for the considered moderate values of~$\tan \beta$. Since~also the coupling~$g_{Hq \bar q}$ is proportional to $\cot \beta$, the contributions from one-loop triangle diagrams involving $H$~exchange are suppressed as well. The~Wilson coefficient $c_N$ in~(\ref{eq:sigmaSI}) is thus dominated by the one-loop triangle contributions that arise from the exchange of the $125 \, {\rm GeV}$ Higgs boson. In the regime $m_A, m_a \gg m_\chi$, we~derive the following approximation
\beq \label{eq:cN}
c_N \simeq -\frac{y_\chi^2}{32 \pi^2} \frac{m_N \hspace{0.5mm} m_\chi}{m_h^2}  \hspace{0.75mm}  f_N \left [ \frac{g_{haa} \sin^2 \theta}{m_a^2}  - \frac{2 \hspace{0.25mm} g_{hAa} \sin \theta \cos \theta}{m_A^2 - m_a^2}  \, \ln \left ( \frac{m_A^2}{m_a^2} \right ) + \frac{g_{hAA} \cos^2 \theta}{m_A^2} \right ] \,, 
\eeq
where $f_N \simeq 0.31$~\cite{Alarcon:2011zs,Alarcon:2012nr,Junnarkar:2013ac,Hoferichter:2015dsa} denotes the relevant effective DM-nucleon coupling. In~the~alignment limit, the trilinear couplings $g_{haa}$, $g_{hAA}$, and $g_{hAa}$ entering~(\ref{eq:cN}) are given by 
\beq \label{eq:ghPPcouplings}
\begin{split}
g_{haa} & = \frac{1}{v^2} \, \Big [ \hspace{0.5mm}  \left (m_h^2 - 2 m_H^2 + 4 m_{H^\pm}^2  - 2 m_a^2  - 2 \lambda_3 v^2 \right ) \sin^2 \theta \\[1mm] & \hspace{1.2cm} - 2 \left (\lambda_{P1} \cos^2 \beta + \lambda_{P2} \sin^2 \beta  \right ) v^2 \cos^2 \theta \hspace{0.5mm} \hspace{0.5mm} \Big ] \,, \\[2mm]
g_{hAa} & = \frac{1}{v^2} \, \Big [ \hspace{0.5mm} m_h^2 - 2 m_H^2 - m_A^2 + 4 m_{H^\pm}^2 - m_a^2 - 2 \lambda_3 v^2 \\[1mm] & \hspace{1.2cm} + 2 \left ( \lambda_{P1} \cos^2 \beta + \lambda_{P2} \sin^2 \beta \right ) v^2 \hspace{0.5mm} \Big ] \sin \theta \cos \theta \,, \\[2mm]
g_{hAA} & = \frac{1}{v^2} \, \Big [ \hspace{0.5mm}  \left (m_h^2 - 2 m_H^2  - 2 m_A^2 + 4 m_{H^\pm}^2  - 2 \lambda_3 v^2 \right ) \cos^2 \theta \\[1mm] & \hspace{1.2cm} - 2 \left ( \lambda_{P1} \cos^2 \beta - \lambda_{P2} \sin^2 \beta \right ) v^2  \sin^2 \theta \hspace{0.5mm} \hspace{0.5mm} \Big ] \,. 
\end{split}
\eeq
Notice that the first, second, and third term in~(\ref{eq:cN}) encodes the effects that stem from loops with a DM particle and two pseudoscalars $a$, one pseudoscalar $A$ and $a$, and two pseudoscalars~$A$, respectively. 

Utilising the formulae~(\ref{eq:cN}) and~(\ref{eq:ghPPcouplings}), we determine that within the parameter space illustrated in the panels of Figure~\ref{fig:exA}, the projected values for the SI DM-nucleon cross section~(\ref{eq:sigmaSI}) can extend up to $\sigma_{\rm SI} = 2.8 \cdot 10^{-47} \, {\rm cm}^2$. The most stringent experimental constraint at $m_\chi = 10 \, {\rm GeV}$ is presently established by the LZ experiment~\cite{LZ:2022lsv}, disfavouring SI DM-nucleon cross sections above $\sigma_{\rm SI} = 1.7 \cdot 10^{-46} \, {\rm cm}^2$ at the $90\%$~CL. These findings indicate that current DM direct detection experiments do not impose constraints on the type-I~2HDM+$a$ benchmarks explored in this study. This finding is consistent with the results presented in~\cite{Arcadi:2024ukq}. Additionally, we note that for $m_\chi = 10 \, {\rm GeV}$, the projected sensitivity of the DARWIN experiment~\cite{DARWIN:2016hyl} to the SI DM-nucleon cross section amount to $\sigma_{\rm SI} = 8.0 \cdot 10^{-48} \, {\rm cm}^2$. Achieving such a remarkable sensitivity would enable the exclusion of triangular regions within the $m_A\hspace{0.125mm}$--$\hspace{0.5mm}m_H$ planes depicted in~Figure~\ref{fig:exA}, where $m_A \lesssim 200 \, {\rm GeV}$ and $m_H \gtrsim 350 \, {\rm GeV}$.

\section{Conclusions}
\label{sec:conclusions}

All previous searches for the 2HDM+$a$ model at the LHC have exclusively considered a Yukawa sector of type~II and mass-degenerate BSM Higgs bosons $A$, $H$, and $H^\pm$. The~assumption about the mass degeneracy, when combined with the constraints from flavour physics that impose lower bounds on the mass of the charged Higgs bosons, implies that the 2HDM bosons are all restricted to be heavier than approximately $600 \, {\rm GeV}$. While these assumptions are well-justified from a phenomenological standpoint, they inherently limit the range of experimental signatures in the 2HDM+$a$ model because they lead to simplified $A$, $H$, and~$H^\pm$ decay patterns. 

In this article, we have presented the first detailed study of 2HDM+$a$ models with a type-I~Yukawa sector. To~be able to define interesting benchmark models, we have provided a comprehensive review of the experimental and theoretical constraints that have to be satisfied to make a given 2HDM+$a$ realisation phenomenologically viable. Our discussion encompasses the constraints arising from SM Higgs physics, flavour observables, LHC searches for BSM Higgs bosons as well as EW precision measurements. Theoretical issues such as vacuum stability and the requirement that the total decay widths of the BSM Higgs bosons are sufficiently small for the NWA to be applicable have also been discussed. The~upshot of our discussion is that in aligned 2HDM+$a$ models with a type-I~Yukawa sector, the constraints on the charged Higgs mass from flavour physics can be lifted for moderate values of $\tan \beta$, allowing the extra Higgs bosons to have masses at or below the EW scale. Additionally, although the constraints from the~$\rho$ parameter are most easily fulfilled for mass-degenerate BSM~Higgs spectra, EW precision measurements currently permit mass splittings between the $H$, $A$, $H^\pm$, and $a$ states of the order of $100 \, {\rm GeV}$ or larger. Motivated by these observations we have given up on the assumption of a mass-degenerate BSM Higgs spectrum. This generalisation has significant implications, notably that previously forbidden decay modes like $A \to H a$, $A\to ZH$, $A \to W^\mp H^\pm$, and others can now be open. By studying the dominant final states that result from $gg \to A$ and $gg \to H$ production followed by the decays of the $A$ and $H$ bosons, we have then identified several interesting $E_{T, \rm miss}$ and non-$E_{T, \rm miss}$ signatures. 

For some of the new signals that have either not been explored or not been interpreted in the 2HDM+$a$ context by ATLAS and CMS, we have devised designated analysis strategies. Specifically, we have performed truth-level analyses for the $h + E_{T, \rm miss}$, $b \bar b + Z$, $b \bar b + E_{T, \rm miss}$, and $ZZ + E_{T, \rm miss}$ signatures. In all instances, the relevant backgrounds and signal yields have been calculated using state-of-the-art theory predictions and MC event generators. Based on our analysis strategies, we have then performed sensitivity studies of the $h + E_{T, \rm miss}$, $b \bar b + Z$, $b \bar b + E_{T, \rm miss}$, and $ZZ + E_{T, \rm miss}$ signals in four distinct type-I~2HDM+$a$ benchmarks. In the resulting 2D~scans of the $m_A\hspace{0.125mm}$--$\hspace{0.5mm}m_H$ plane, we have also depicted the constraints that arise from searches for light charged Higgs bosons, EW precision measurements and indicated theoretical restrictions that for example follow from vacuum-stability arguments. The 2D scans shown in~Figure~\ref{fig:exA} combine all discussed constraints and represent the main result of our work. Our sensitivity studies suggest that the examined~$E_{T, \rm miss}$ and non-$E_{T, \rm miss}$ signatures allow to probe complementary regions of parameter space of type-I~2HDM+$a$ realisations of the form~(\ref{eq:generalparameter}). From the panels in~Figure~\ref{fig:exA}, it is however also evident that the analysed constraints from $h + E_{T, \rm miss}$, $b\bar b + Z$, $b \bar b + E_{T, \rm miss}$, $ZZ + E_{T, \rm miss}$, and charged Higgs searches are not sufficient to probe the entire $m_A\hspace{0.125mm}$--$\hspace{0.5mm}m_H$ planes. Processes that could offer sensitivity to the unexplored parameter space depicted in~Figure~\ref{fig:exA} include $Z + E_{T, \rm miss}$ production~\cite{ATLAS:2019wdu,CMS:2020ulv,ATLAS:2021gcn}, and with the anticipated large dataset at the HL-LHC, also the $t \bar t$ \cite{ATLAS:2017snw,CMS:2019pzc,ATLAS-CONF-2024-001} and $4t$ \cite{CMS:2019rvj,ATLAS:2020hpj,ATLAS:2021kqb,CMS:2023zdh,ATLAS:2023ajo} channels. These processes have, however, not been analysed in detail in this article. We have finally pointed out that in the context of the type-I~2HDM+$a$ benchmark models featured in~Figure~\ref{fig:muH}, $hh + E_{T, \rm miss}$ signatures are predicted in parts of the parameter space that should be detectable at the~HL-LHC. A~comprehensive study of~the $hh + E_{T, \rm miss}$ channel in the 2HDM+$a$ model therefore seems worthwhile, but is beyond the~scope of the current work. 

An alternative approach to extend the sensitivity towards lower values of the mass of the pseudoscalar $A$ is to develop searches for the $h+E_{T,\rm miss}$ or $b\bar{b}+E_{T,\rm miss}$ final states utilising $b$-jet triggers~\cite{ATLAS:2021piz}. Similar investigations targeting the production of light DM mediators~\cite{ATLAS:2022ygn} have demonstrated the viability of $b$-jet triggers in probing signals with low~$E_{T,\rm miss}$. Therefore, incorporating such triggers in updated versions of~\cite{ATLAS:2021shl,ATLAS:2023zkt} would likely expand the parameter space probed by both the $h+E_{T,\rm miss}$ and $b\bar{b}+E_{T,\rm miss}$ signatures depicted in~Figure~\ref{fig:exA}. Similar considerations apply to $ZZ + E_{T, \rm miss}$ production, where existing searches~\cite{ATLAS:2021wob,ATLAS:2024bzr} utilise the four-lepton invariant mass as a discriminant. As~demonstrated in our study, employing the transverse mass constructed from the $4 \ell$ system and $E_{T, \rm miss}$ would provide an improved signal-to-background separation in this case. Finally, we note that a potentially detectable $gg \to A \to W^\mp H^\pm \to W^\mp tb$ signal can emerge within the type-I 2HDM+$a$ model. In the framework of the pure 2HDM model of type I featuring a light scalar $H$, the significance of resonant $gg \to A \to W^\mp H^\pm$ production has been highlighted in~\cite{Haisch:2017gql}. This production mechanism allows for novel LHC signatures for which, to the best of our knowledge, no dedicated searches by ATLAS or CMS currently exist, thus warranting further investigations.

\acknowledgments We acknowledge helpful exchanges with Susanne Westhoff and Jose Zurita concerning~\cite{Blanke:2019hpe}. A big thank you also goes to Thomas Hahn and Luc Schnell for computer support. IK~furthermore thanks the Max-Planck-Institut f{\"u}r Physik for their hospitality during the initial phase of this work. SA and IK are funded by the German Research Foundation (DFG) under grant No.~AR~1321/1-1. We are grateful to Daniel Puigmart\'i~Molano for pointing out the issues with the BFB conditions discussed in~Section~\ref{sec:theoryconstraints}.

\begin{appendix}

\section[Impact of $\bm{gg \to h 4\chi}$ contribution to the $\bm{h+E_{T, \rm miss}}$ signal]{Impact of $\bm{gg \to h 4\chi}$ contribution to the $\bm{h+E_{T, \rm miss}}$ signal}
\label{app:h4chi}

In the type-I~2HDM+$a$ model featuring non-degenerate BSM Higgs bosons, the $h + E_{T, \rm miss}$ signature typically involves contributions from both $gg \to h \chi \bar \chi$ and $gg \to h 4 \chi$ production. Sample diagrams are displayed on the left in~Figure \ref{fig:diagrams1} and on the left in~Figure \ref{fig:diagrams6}. To~precisely determine the LHC sensitivity of $h + E_{T, \rm miss}$ search strategies such as the one detailed in~Section~\ref{sec:hMETsearch}, one should hence include both channels. While the event generation for the $gg \to h \chi \bar \chi$ contribution poses no issue, generating event samples for $gg \to h 4 \chi$ is quite laborious due to the large number of contributing diagrams in the latter case (cf.~Appendix~\ref{app:hMETsignal}). Given~this challenge, our sensitivity study of the $h+E_{T, \rm miss}$ signal presented in Section~\ref{sec:constraintssummary}, relies solely on $gg \to h \chi \bar \chi$ signal samples.

\begin{figure}[t!]
\begin{center}
\includegraphics[width=0.725\textwidth]{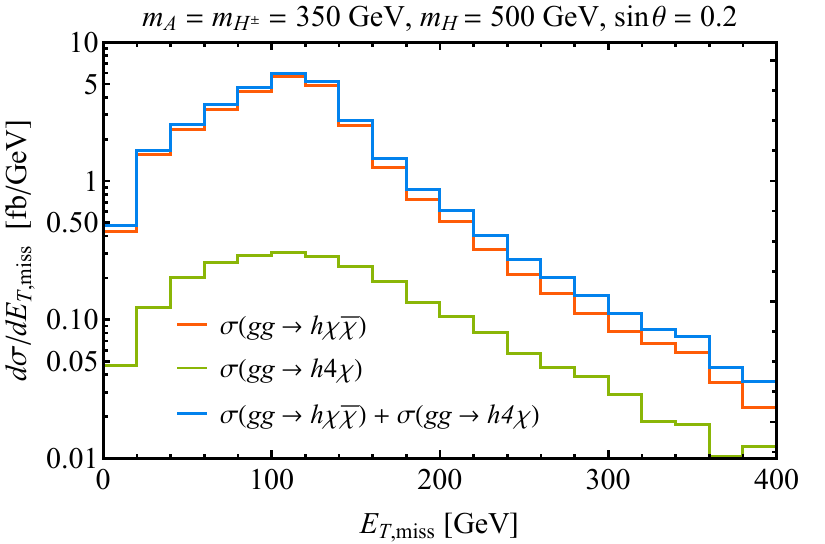}
\vspace{2mm}
\caption{\label{fig:hMET_MET} Predictions for the $E_{T, \rm miss}$ spectrum in $gg \to h \chi \bar \chi$ and $gg \to h 4 \chi$ production, as well as their combination. The chosen values for the masses~$m_A$, $m_H$, and $m_{H^\pm}$, along with the sine of the pseudoscalar mixing angle $\theta$, are specified in the panel heading. The~remaining parameters are set according to (\ref{eq:generalparameter}). Additional explanations can be found in the main~text.} 
\end{center}
\end{figure}

To assess the potential impact of the $gg \to h 4\chi$ contribution on the $h+E_{T, \rm miss}$ signal, we depict the $E_{T, \rm miss}$ spectrum of $gg \to h \chi \bar \chi$ and $gg \to h 4 \chi$ for the same parameter set in~Figure~\ref{fig:hMET_MET}. The results are presented without any experimental cuts. The parameter selections are $m_A = m_{H^\pm} = 350 \, {\rm GeV}$, $m_H = 500 \, {\rm GeV}$, $\sin \theta = 0.2$, along with~(\ref{eq:generalparameter}). As~shown in~Figure~\ref{fig:exA}, at LHC~Run~2, the chosen $m_A$ and $m_H$ values cannot be excluded by our truth-level analysis of the $h+E_{T, {\rm miss}}$ signature, which is based on $gg \to h \chi \bar{\chi}$ samples only. The~inclusion of the $gg \to h 4\chi$ channel, which has a somewhat harder $E_{T, {\rm miss}}$ spectrum compared to $gg \to h \chi \bar \chi$, in regions where the $H\to Aa$ decay is kinematically allowed, would slightly enhance the sensitivity of our search strategy. For~the parameter choices in~Figure~\ref{fig:hMET_MET}, we find that the inclusion of the $gg \to h 4\chi$ contribution leads to an enhancement of the $h + E_{T, {\rm miss}}$ cross section by around~$25\%$ in the fiducial region with $E_{T, {\rm miss}} > 150 \, {\rm GeV}$. Notice that there is no interference between the $gg \to h \chi \bar \chi$ and $gg \to h 4 \chi$ channels, meaning that the event samples can simply be added to obtain the full~$h + E_{T, {\rm miss}}$ yield. We finally add that for the parameter choices made in Figure~\ref{fig:hMET_MET}, almost the entire contribution to $gg \to h 4 \chi$ production arises from the channel $gg \to h aa \to h 4 \chi$. This observation applies to both the inclusive cross section and the $E_{T, \text{miss}}$ spectrum, which are found to be indistinguishable within the statistical uncertainties of our MC simulations.

\section{Details on the signal generation}
\label{app:signal}

The sensitivity studies presented in this work are all based on MC simulations that use an {\tt UFO} implementation of the type-I~2HDM+$a$~model as described in Section~\ref{sec:2hdma}. Our~new~{\tt UFO} is called {\tt Pseudoscalar\_2HDMI} and is provided at~\cite{UFO}. After starting {\tt MadGraph5\_aMC@NLO}, the {\tt UFO} model is loaded with:
\begin{verbatim}
     MG5_aMC> import model Pseudoscalar_2HDMI
\end{verbatim}
This command restricts the {\tt Pseudoscalar\_2HDMI} model using the restriction card file {\tt models/Pseudoscalar\_2HDMI/restrict\_default.dat}. Additional restriction cards are included alongside our {\tt UFO} implementation, which can be utilised to streamline event generation. For example, signals in the aligned type-I~2HDM+$a$ model can be generated by employing the restriction card file {\tt models/Pseudoscalar\_2HDMI/restrict\_aligned.dat}. This~is achieved by first loading the relevant model via 
\begin{verbatim}
     MG5_aMC> import model Pseudoscalar_2HDMI-aligned
\end{verbatim}
and then generating the process of interest. The generation is simplified since Feynman diagrams that involve couplings that vanish for $\cos \left ( \beta - \alpha \right ) = 0$ are automatically discarded in the calculation of the matrix elements, leading in general to a reduced number of contributing graphs. 

Below we will give details on the generation of the $h + E_{T, \rm miss}$, $b \bar b + Z$, $b \bar b + E_{T, \rm miss}$, $ZZ + E_{T, \rm miss}$, and $hh + E_{T, \rm miss}$ signatures. Other channels can be generated in a similar fashion. The used commands assume that the {\tt Pseudoscalar\_2HDMI-aligned} model has been loaded by invoking the relevant {\tt MadGraph5\_aMC@NLO} command.

\subsection[$h + E_{T, \rm miss}$ sample]{$\bm{h + E_{T, \rm miss}}$ sample}
\label{app:hMETsignal}

The {\tt MadGraph5\_aMC@NLO} syntax used in Section~\ref{sec:hMETsearch} to generate the $gg \to h\chi \bar \chi$ contribution to the~$h + E_{T, \rm miss}$ signal is 
\begin{verbatim}
     MG5_aMC> generate g g > h1 xd xd~ [noborn=QCD]
\end{verbatim}
where {\tt [noborn=QCD]} indicates that one deals with a loop-induced process. Furthermore, \verb+h1+, \verb+xd+, and \verb+xd~+ denote the $125 \, {\rm GeV}$ Higgs boson $h$, the DM candidate $\chi$ and its anti-particle~$\bar \chi$, respectively. Notice that the generated matrix elements include not only the resonant $gg \to A \to h a$ contribution illustrated on the left in~Figure~\ref{fig:diagrams1}, but also various non-resonant diagrams, such as the contribution depicted on the right-hand side in the same figure. This results in a total of 56 contributing graphs. The decay of the Higgs into~$b \bar b$ pairs,~i.e.~$h \to b \bar b$, is performed with {\tt MadSpin} in our signal generation. 

The full $gg \to h 4 \chi$ contribution to the~$h + E_{T, \rm miss}$ signature can be generated via
\begin{verbatim}
     MG5_aMC> generate g g > h1 xd xd~ xd xd~ [noborn=QCD]
\end{verbatim}
This process contains 3776 Feynman diagrams. One of the relevant graphs is depicted on the right-hand side in~Figure~\ref{fig:diagrams6}. We would like to caution the reader that due to the large number of diagrams involved, generating events for the exact $gg \to h 4 \chi$ contribution requires significant computing resources. The~$125 \, {\rm GeV}$ Higgs boson can be decayed into~$b \bar b$  pairs using {\tt MadSpin}. 

The computation of the full $gg \to h 4 \chi$ contribution to the $h + E_{T, \text{miss}}$ signature can be significantly simplified by generating the process
\begin{verbatim}
     MG5_aMC> generate g g > h1 h4 h4 [noborn=QCD]
\end{verbatim}
and simulating the decays $h \to b \bar b$ and $a \to \chi \bar \chi$ using \texttt{MadSpin}. Since this process contains only 296 Feynman diagrams compared to the 3776 graphs relevant for the full computation of $gg \to h 4 \chi$, the event generation of $gg \to h aa$ poses no issue. For the parameter choices made in Appendix~\ref{app:h4chi}, the kinematic distributions of the full $gg \to h 4 \chi$ process are essentially indistinguishable from those of the $gg \to h aa \to h 4 \chi$ transition.

\subsection[$b \bar b + Z$ sample]{$\bm{b \bar b + Z}$ sample}
\label{app:bbZsignal}

The $b \bar b + Z$ analysis strategy described in~Section~\ref{sec:ZHsearches}, relies on signal samples generated using the following command: 
\begin{verbatim}
     MG5_aMC> generate g g > z h2 [noborn=QCD]
\end{verbatim}
Here \verb+z+ denotes the $Z$ boson, while \verb+h2+ is the $H$ boson. In total 20 diagrams are generated, including the resonant $gg \to A\to ZH$ graph shown on the right in~Figure~\ref{fig:diagrams4}. The decays $Z \to \ell^+ \ell^-$ and $H \to b \bar b$ are handled by {\tt MadSpin}, where spin correlations are neglected due to limitations in the {\tt MadSpin} code~\cite{Hirschi:2015iia}.

\subsection[$b \bar b + E_{T, \rm miss}$ sample]{$\bm{b \bar b + E_{T, \rm miss}}$ sample}
\label{app:bbMETsignal}

Since the $b \bar b + E_{T, \rm miss}$ signature, where the $b \bar b$ pair originates from the decay $H \to b \bar b$, receives contributions from both $gg \to Ha \to b \bar b \chi \bar \chi$ and $gg \to HZ \to b \bar b \nu \bar \nu$ production, generating the corresponding event samples is a bit more involved. We have used 
\begin{verbatim}
     MG5_aMC> generate g g > h2 xd xd~ [noborn=QCD]
     MG5_aMC> add process g g > z h2 [noborn=QCD]
\end{verbatim}
resulting in a total number 76 of resonant and non-resonant diagrams. Example graphs are on the left in~Figure~\ref{fig:diagrams5} and on the right in~Figure~\ref{fig:diagrams4}, respectively. The decays $H \to b \bar b$ and $Z \to \nu \bar \nu $ are performed with {\tt MadSpin}, neglecting spin correlations. The event generation described above yields the signal samples targeted by the analysis strategy outlined in~Section~\ref{sec:HMETsearches}.

\subsection[$Z Z + E_{T, \rm miss}$ sample]{$\bm{Z Z + E_{T, \rm miss}}$ sample}
\label{app:ZZMETsignal}

In the case of the $Z Z + E_{T, \rm miss}$ signal studied in Section~\ref{sec:ZZMETsearches}, the {\tt MadGraph5\_aMC@NLO} syntax to generate the corresponding matrix elements reads:
\begin{verbatim}
     MG5_aMC> generate g g > z z xd xd~ [noborn=QCD]
\end{verbatim}
This command generates a total of 248 different Feynman diagrams, which include the resonant contribution $gg \to A \to ZH \to ZZa$ depicted on the right-hand side of~Figure~\ref{fig:diagrams5}. The $Z \to \ell^+ \ell^-$ decays are simulated with {\tt MadSpin}. Spin-correlation effects are neglected in this simulation. 

\subsection[$hh + E_{T, \rm miss}$ sample]{$\bm{hh + E_{T, \rm miss}}$ sample}
\label{app:hhMETsignal}

The syntax in {\tt MadGraph5\_aMC@NLO} to generate the complete $hh + E_{T, \rm miss}$ signature, which includes the resonant diagram depicted on the right-hand side of~Figure~\ref{fig:diagrams8}, is as follows:
\begin{verbatim}
     MG5_aMC> generate g g > h1 h1 xd xd~ [noborn=QCD]
\end{verbatim}
The matrix element generated in this manner comprises a total of 616 Feynman diagrams. The decay of the $125 \, {\rm GeV}$ Higgs boson can then, for example, be simulated using {\tt MadSpin}. Notice that the sensitivity estimate presented~in~Section~\ref{sec:others} is not based on the generation of a full $hh + E_{T, \rm miss}$ sample. Instead, it only takes into account the contribution from the resonant $gg \to H \to AA \to haha \to 4b 4\chi$ channel.
\end{appendix}



%

\end{document}